\documentclass{aa}

\usepackage{wasysym}

\usepackage{graphicx}
\usepackage{amsmath}   
\usepackage{tabulary}
\usepackage{epsfig}
\usepackage{amssymb}
\usepackage{multirow}
\usepackage{appendix}
\usepackage{natbib}
\usepackage{lineno}
\usepackage{wasysym}
\usepackage{color}         
\usepackage{enumerate}
\usepackage{tikz}

\usepackage{graphicx,url,twoopt,natbib}
\usepackage{txfonts}
\usepackage[pdftitle={Predictable patterns in planetary transit timing variations and transit duration variations due to exomoons}, colorlinks = true, breaklinks = true, citecolor = blue, linkcolor = blue, urlcolor = blue, pdfauthor = {Ren\'{e} Heller}]{hyperref}
\bibpunct{(}{)}{;}{a}{}{,}    

\begin{document} 

\title{Predictable patterns in planetary transit timing variations and transit duration variations due to exomoons}

\titlerunning{Predictable patterns in planetary TTVs and TDVs due to exomoons}

\author{Ren\'{e} Heller\inst{1}
             \and
             Michael Hippke\inst{2}
             \and
             Ben Placek\inst{3}
             \and
             Daniel Angerhausen\inst{4,5}
             \and
             Eric Agol\inst{6,7}
}

\institute{Max Planck Institute for Solar System Research, Justus-von-Liebig-Weg 3, 37077 G\"ottingen, Germany; \href{mailto:heller@mps.mpg.de}{heller@mps.mpg.de}
               \and
               Luiter Stra{\ss}e 21b, 47506 Neukirchen-Vluyn, Germany; \href{mailto:hippke@ifda.eu}{hippke@ifda.eu}
               \and
               Center for Science and Technology, Schenectady County Community College, Schenectady, NY 12305, USA;\\
               \href{placekbh@sunysccc.edu}{placekbh@sunysccc.edu}
               \and
               NASA Goddard Space Flight Center, Greenbelt, MD 20771, USA; \href{mailto:daniel.angerhausen@nasa.gov}{daniel.angerhausen@nasa.gov}
               \and
               USRA NASA Postdoctoral Program Fellow, NASA Goddard Space Flight Center, 8800 Greenbelt Road, Greenbelt, MD 20771, USA
               \and
               Astronomy Department, University of Washington, Seattle, WA 98195, USA; \href{mailto:agol@uw.edu}{agol@uw.edu}
               \and
               NASA Astrobiology Institute's Virtual Planetary Laboratory, Seattle, WA 98195, USA
}

\date{Received 22 March 2016; Accepted 12 April 2016}
 
\abstract{
We present new ways to identify single and multiple moons around extrasolar planets using planetary transit timing variations (TTVs) and transit duration variations (TDVs). For planets with one moon, measurements from successive transits exhibit a hitherto undescribed pattern in the TTV-TDV diagram, originating from the stroboscopic sampling of the planet's orbit around the planet--moon barycenter. This pattern is fully determined and analytically predictable after three consecutive transits. The more measurements become available, the more the TTV-TDV diagram approaches an ellipse. For planets with multi-moons in orbital mean motion resonance (MMR), like the Galilean moon system, the pattern is much more complex and addressed numerically in this report. Exomoons in MMR can also form closed, predictable TTV-TDV figures, as long as the drift of the moons' pericenters is sufficiently slow. We find that MMR exomoons produce loops in the TTV-TDV diagram and that the number of these loops is equal to the order of the MMR, or the largest integer in the MMR ratio. We use a Bayesian model and Monte Carlo simulations to test the discoverability of exomoons using TTV-TDV diagrams with current and near-future technology. In a blind test, two of us (BP, DA) successfully retrieved a large moon from simulated TTV-TDV by co-authors MH and RH, which resembled data from a known \textit{Kepler} planet candidate. Single exomoons with a 10\,\% moon-to-planet mass ratio, like to Pluto-Charon binary, can be detectable in the archival data of the \textit{Kepler} primary mission. Multi-exomoon systems, however, require either larger telescopes or brighter target stars. Complementary detection methods invoking a moon's own photometric transit or its orbital sampling effect can be used for validation or falsification. A combination of \textit{TESS}, \textit{CHEOPS}, and \textit{PLATO} data would offer a compelling opportunity for an exomoon discovery around a bright star.
}


\maketitle

\section{Introduction}
\label{sec:introduction}

The search for moons around planets beyond the solar system is entering a critical phase. The first dedicated exomoon surveys have now been implemented using space-based highly accurate \textit{Kepler} photometry \citep{2012ApJ...750..115K,2013A&A...553A..17S,2015ApJ...806...51H} and more will follow in the near future using \textit{CHEOPS} \citep{2015PASP..127.1084S} and \textit{PLATO} \citep{2015ApJ...810...29H}. An important outcome of the first exomoon searches is that moons at least twice as massive as Ganymede, the most massive local moon, are rare around super-Earths \citep{2015ApJ...813...14K}.

More than a dozen techniques have been proposed to search for exomoons: 
(i.) transit timing variations \citep[TTVs;][]{1999A&AS..134..553S,2007A&A...470..727S} and transit duration variations \citep[TDVs;][]{2009MNRAS.392..181K,2009MNRAS.396.1797K}\,of exoplanets;
(ii.)\,the direct photometric transit signature of exomoons \citep{1999A&AS..134..553S,2001ApJ...552..699B,2006A&A...450..395S,2006ApJ...636..445C,2007A&A...476.1347P,2011ApJ...743...97T,2011MNRAS.416..689K};
(iii.) microlensing \citep{2002ApJ...580..490H,2010A&A...520A..68L,2014ApJ...785..155B};
(iv.) mutual eclipses of directly imaged, unresolved planet--moon binaries \citep{2007A&A...464.1133C};
(v.) the wobble of the photometric center of unresolved, directly imaged planet--moon systems \citep{2007A&A...464.1133C,2015ApJ...812....5A}
(vi.) time-arrival analyses of planet--moon systems around pulsars \citep{2008ApJ...685L.153L};
(vii.) planet--moon mutual eclipses during stellar transits \citep{2009PASJ...61L..29S,2012MNRAS.420.1630P};
(viii.) the Rossiter-McLaughlin effect \citep{2010MNRAS.406.2038S,2012ApJ...758..111Z};
(ix.) scatter peak analyses \citep{2012MNRAS.419..164S} and the orbital sampling effect of phase-folded light curves\,\citep{2014ApJ...787...14H,2016HHJ};
(x.) modulated radio emission from giant planets with moons  \citep{2014ApJ...791...25N,2016ApJ...821...97N};
(xi.) the photometric detection of moon-induced plasma torii around exoplanets \citep{2014ApJ...785L..30B};
(xii.) and several other spectral \citep{2004AsBio...4..400W,2011ApJ...741...51R,2014ApJ...796L...1H} and photometric \citep{2009AsBio...9..269M,2013ApJ...769...98P} analyses of the infrared light in exoplanet-exomoon systems.\footnote{Naturally, these methods are not fully independent, and our enumeration is somewhat arbitrary.}

None of these techniques has delivered a secure exomoon detection as of today, which is partly because most of these methods are reserved for future observational technologies. Some methods are applicable to the available data, for example, from the \textit{Kepler} primary mission, but they are extremely computer intense \citep{2012ApJ...750..115K}. We present a novel method to find and characterize exomoons that can be used with current technologies and even the publicly available \textit{Kepler} data. We identify a new pattern in the TTV-TDV diagram of exoplanet-exomoon systems that allows us to distinguish between single and multiple exomoons. Detection of multiple moons is naturally more challenging than the detection of single exomoons owing to the higher complexity of models that involve multiple moons \citep{2014ApJ...787...14H,2015ApJ...810...29H,2015ApJ...813...14K,2016HHJ,2016ApJ...821...97N}.

\section{Patterns in the TTV-TDV diagram}
\label{sec:patters}

An exoplanet transiting a star can show TTVs \citep{1999A&AS..134..553S} and TDVs \citep{2009MNRAS.392..181K,2009MNRAS.396.1797K} if accompanied by a moon. Various ways exist to measure TTVs, for example, with respect to the planet-moon barycenter or photocenter \citep{2006A&A...450..395S,2007A&A...470..727S,2015PASP..127.1084S}. We utilize the barycentric TTV. On their own, TTVs and TDVs yield degenerate solutions for the satellite mass ($M_{\rm s}$) and the orbital semimajor axis of the satellite around the planet ($a_{\rm s}$). If both TTV and TDV can be measured repeatedly, however, and sources other than moons be excluded, then the $M_{\rm s}$ and $a_{\rm s}$ root mean square (RMS) values can be estimated and the degeneracy be solved \citep{2009MNRAS.392..181K,2009MNRAS.396.1797K}.

This method, however, works only for planets (of mass $M_{\rm p}$) with a single moon. Moreover, observations will always undersample the orbit of the moon and $P_{\rm s}$ cannot be directly measured \citep{2009MNRAS.392..181K}. This is because the orbital period of the planet around the star ($P_{\rm p}$) is $\gtrsim9$ times the orbital period of the satellite around the planet ($P_{\rm s}$) to ensure Hill stability. Strictly speaking, it is the circumstellar orbital period of the planet--moon barycenter ($P_{\rm B}$) that needs to be ${\gtrsim}~9\,P_{\rm s}$, but for $M_{\rm s}/M_{\rm p}~{\rightarrow}~0$ we have $P_{\rm p}~{\rightarrow}~P_{\rm B}$. Finally, a predictable pattern in TTV-TDV measurements has not been published to date.

We present new means to determine (1) the remainder of the division of the orbital periods of the moon and the planet for one-moon systems; (2) the TTV and TDV of the planet during the next transit for one-moon systems; and (3) the number of moons in multiple moon systems.

Our approach makes use of the fact that TTVs and TDVs are phase-shifted by $\pi/2$ radians, as first pointed out by \citet{2009MNRAS.392..181K}. In comparison to the work of Kipping, however, we fold the TTV-TDV information into one diagram rather than treating this information as separate functions of time.

\subsection{Exoplanets with a single exomoon}
\label{sec:single}

Imagine a transiting exoplanet with a single moon. We let $P_{\rm B}$ be $f~=~n+r$ times $P_{\rm s}$, where $n$ is an integer and $0~{\leq}~r~{\leq}~1$ is the remainder. In the Sun-Earth-Moon system, for example, we have $P_{\rm B}~\approx~365.25$\,d, $P_{\rm s}~\approx~27.3$\,d, and thus $P_{\rm B}~\approx~13.39\,P_{\rm s}$. Hence, $n~=~13$ and $r~\approx~0.39$.

We measure the transit timing and transit duration of the first transit in a series of transit observations. These data do not deliver any TTV or TDV, because there is no reference value yet to compare our measurements to yet. After the second transit, we may obtain a TDV because the duration of the transit might be different from the first transit.  We are still not able to obtain a TTV because determining the transit period requires two consecutive transits to be observed. Only after the third transit are we able to measure a variation of the transit period, i.e., our first TTV (and our second TDV). While we observe more and more transits, our average transit period and transit duration values change and converge to a value that is unknown a priori but that could be calculated analytically for one-moon systems if the system properties were known.

\begin{figure}[!h]
 \centering
  \includegraphics[angle= 0, width=0.487\linewidth]{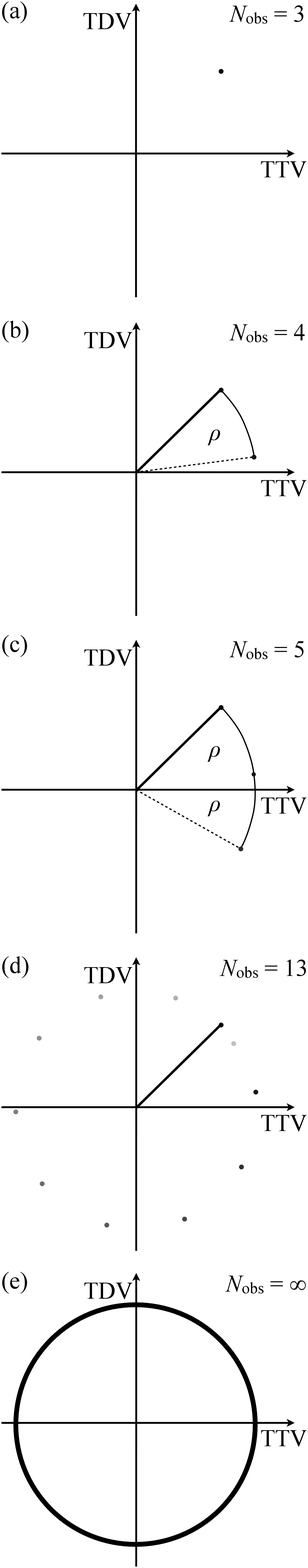}
  \caption{Evolution of a TTV-TDV diagram of an exoplanet with 1 moon. TTV and TDV amplitudes are normalized to yield a circular figure.}
  \label{fig:method}
 \end{figure}

Figure~\ref{fig:method} qualitatively shows such an evolution of a TTV-TDV diagram. Different from the above-mentioned procedure, measurements in each panel are arranged in a way to gradually form the same figure as shown in panel (e), as if the average TTV and TDV for $N_{\rm obs}=\infty$ were known a priori. In each panel, $N_{\rm obs}$ denotes the number of consecutive transit observations required to present those hypothetical data points. In panels (b)-(d), the first data point of the data series is indicated by a solid line. Generally, TTV and TDV amplitudes are very different from each other, so the final figure of the TTV-TDV diagram in panel (e) is an ellipse rather than a circle. However, if the figure is normalized to those amplitudes and if the orbits are circular, then the final figure is a circle as shown. The angle $\rho$ corresponds to the remainder of the planet-to-moon orbital period ratio, which can be deduced via $r=\rho/(2\pi)$. Knowledge of $\rho$ or $r$ makes it possible to predict the planet-moon orbital geometry during stellar transits, which enables dedicated observations of the maximum planet-moon apparent separation (at TTV maximum values) or possible planet-moon mutual eclipses (at TDV maximum values). Yet, $n$ remains unknown and it is not possible to determine how many orbits the moon has completed around the planet during one circumstellar orbit of their common barycenter. As an aside, $\rho$ cannot be used to determine the sense of orbital motion of the planet or moon \citep{2014ApJ...791L..26L,2014ApJ...796L...1H}.

If the moon's orbit is in an $n:1$ orbital mean motion resonance (MMR) with the circumstellar orbit, then the moon always appears at the same position relative to the planet during subsequent transits. Hence, if $\rho=0=r$, there is effectively no TTV or TDV and the moon remains undiscovered. In an $(n+\frac{1}{2}):1$ MMR, we obtain $r=\frac{1}{2}$ and subsequent measurements in the TTV-TDV diagram jump between two points. Again, the full TTV-TDV figure is not sampled and the moon cannot be characterized. In general, in an $(n+\frac{1}{x}):1$ MMR, the diagram jumps between $x$ points and $r=\frac{1}{x}$.

\begin{figure}[!h]
\includegraphics[width=.796\linewidth, angle=-90]{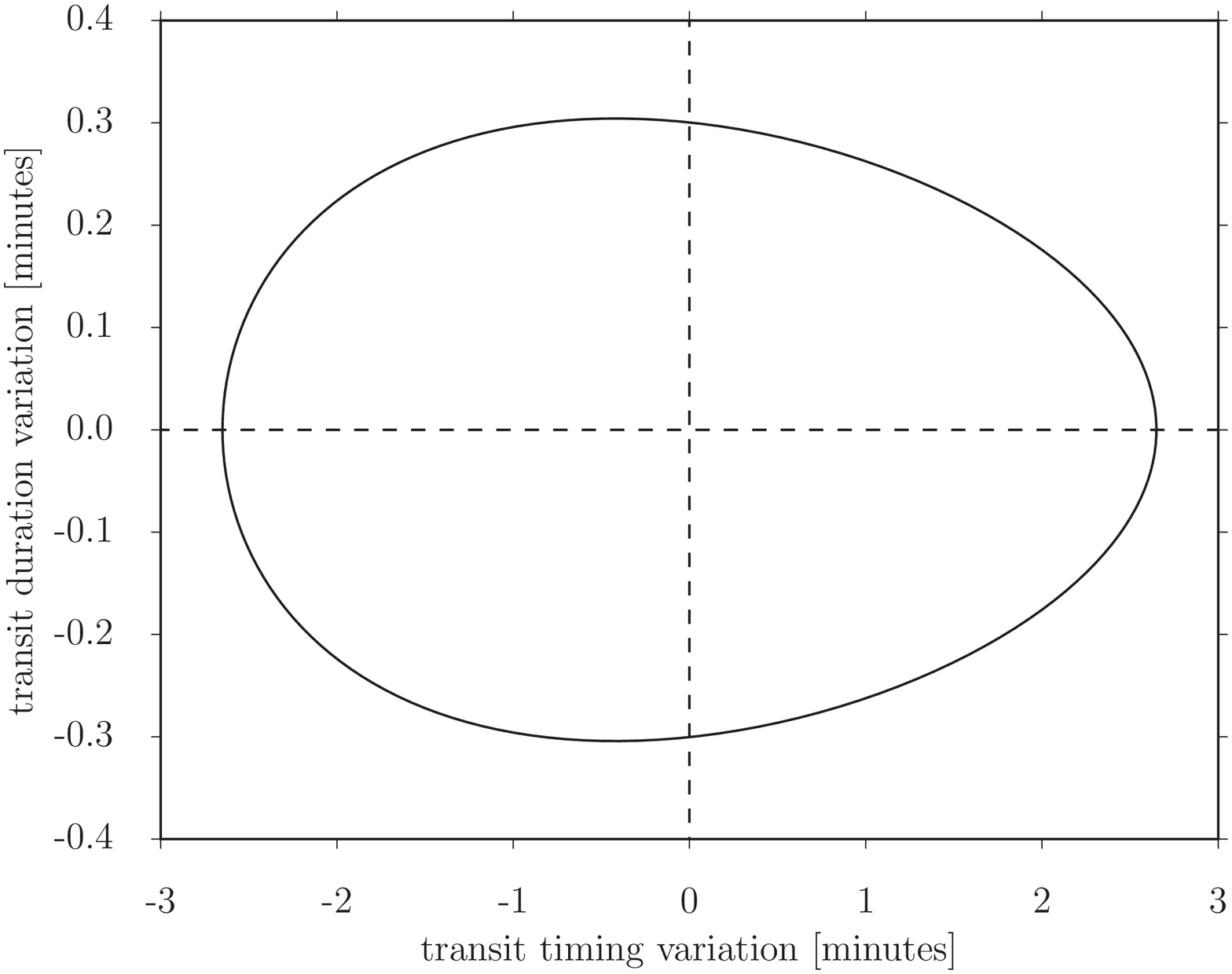}
\caption{\label{fig:ecc} TTV-TDV diagram for the Earth-Moon system transiting the Sun, but with the orbital eccentricity of the Moon increased to 0.25.}
\end{figure}

Orbital eccentricities of the satellite ($e_{\rm s}$) complicate this picture. Figure~\ref{fig:ecc} shows the TTV-TDV diagram for a planet--moon system akin to the Earth-Moon binary around a Sun-like star. The only difference that we introduced is an $e_{\rm s}$ value of $0.25$. The resulting shape of the TTV-TDV can be described as an egg-shaped ellipsoid with only one symmetry axis that cannot be transformed into an ellipse by linear scaling of the axis. The orientation of that figure depends on the orientation of the periastron of the planet--moon system.

\subsection{Exoplanets with multiple exomoons}

For single moons on circular orbits, the TTV-TDV diagram can be calculated analytically if the TTV and TDV amplitudes are known \citep{1999A&AS..134..553S,2009MNRAS.392..181K,2009MNRAS.396.1797K}. For systems with more than $N=2$ bodies, however, analytical solutions do not exist. Hence, for systems with more than one moon (and one planet), we resort to numerical simulations.

\subsubsection{From $N$-body simulations to TTVs and TDVs}
\label{subsub:nbody}

In the following, we generate TTV-TDV diagrams of exoplanet-exomoons systems using a self-made, standard $N$-body integrator that calculates the Newtonian gravitational accelerations acting on $N$ point masses.

The average orbital speeds of the major solar system moons are well known.\footnote{\href{http://nssdc.gsfc.nasa.gov/planetary/factsheet/joviansatfact.html}{http://nssdc.gsfc.nasa.gov/planetary/factsheet/joviansatfact.html}} The initial planetary velocities in our simulations, however, are unknown and need to be calculated. We take $v_{\rm B}=0$ as the velocity of the planet--moon barycenter, assuming that the system is unperturbed by other planets or by the star. This is a reasonable assumption for planets at about 1\,au from a Sun-like star and without a nearby giant planet. In the Earth-Moon system, for reference, perturbations from mostly Venus and Jupiter impose a forced eccentricity of just 0.0549 \citep{2007Sci...318..244C}, which is undetectable for exomoon systems in the foreseeable future. We let the total mass of the planet--moon system, or the mass of its barycenter, be $m_{\rm B}$. With index p referring to the planet, and indices s1, s2, etc. referring to the satellites, we have

\begin{align}\label{eq:barycenter} \nonumber
v_{\rm B} &= \frac{v_{\rm p} m_{\rm p} + v_{\rm s1} m_{\rm s1} + v_{\rm s2} m_{\rm s2} + \dots}{m_{\rm B}} = 0 \\ \nonumber \\
\Leftrightarrow v_{\rm p} &= \frac{-v_{\rm s1} m_{\rm s1} - v_{\rm s2} m_{\rm s2} - \dots}{m_{\rm p}}
\end{align}

Our simulations run in fixed time steps. We find that $10^3$ steps per orbit are sufficient to keep errors in TTV and TDV below $0.1$\,s in almost all cases. In cases with very small moon-to-planet mass ratios, we used $10^4$ steps to obtain $<0.1$\,s errors. A TTV/TDV error of 0.1\,s would generally be considered to be too large for studies dedicated to the long-term orbital evolution and stability of multiplanet systems. For our purpose of simulating merely a single orbit to generate the corresponding TTV-TDV diagram, however, this error is sufficient because the resulting errors in TTV and TDV are smaller than the linewidth in our plots. The processing runtime for $10^4 $ steps in a five-moon system is $<1$\,s on a standard, commercial laptop computer. Our computer code, used to generate all of the following TTV-TDV figures, is publicly available with examples under an open source license.\footnote{\href{https://github.com/hippke/TTV-TDV-exomoons}{https://github.com/hippke/TTV-TDV-exomoons}}

Under the assumption that the orbits of the planet--moon barycenter around the star and of the planet with moons around their local barycenter are coplanar, TTVs depend solely on the variation of the sky-projected position of the planet relative to the barycenter. On the other hand, TDVs depend solely on the variation of the orbital velocity component of the planet that is tangential to the line of sight.\footnote{In case of nonaligned orbits, the planet shows an additional TDV component \citep{2009MNRAS.396.1797K}.} For a given semimajor axis of the planet--moon barycenter ($a_{\rm B}$) and $P_{\rm B}$ around the star, we first compute the orbital velocity of the planet on a circular circumstellar orbit. We then take the stellar radius ($R_\star$) and convert variations in the relative barycentric position of the planet into TTVs and variations in its orbital velocity into TDVs. Each of these $\approx10^3$ orbital configurations corresponds to one stellar transit of the planet--moon system. It is assumed that the planet--moon system transits across the stellar diameter, that is, with a transit impact parameter of zero, but our general conclusions would not be affected if this condition were lifted. In the Sun--Earth--Moon system, for example, Earth's maximum tangential displacement of 4\,763\,km from the Earth--Moon barycenter corresponds to a TTV of 159\,s compared to a transit duration of about 13\,hr. As we are interested in individual, consecutive TTV and TDV measurements, we use amplitudes rather than RMS values.

Our assumption of coplanar orbits is mainly for visualization purposes, but more complex configurations with inclined orbits should be revisited in future studies to investigate the effects of variations in the planetary transit impact parameter \citep{2009MNRAS.396.1797K} and the like. For this study, coplanar cases can be justified by the rareness of high moon inclinations ($i_{\rm s}$) in the solar system; among the 16 largest solar system moons, only four have inclinations $>1^{\circ}$; Moon (5.5$^{\circ}$), Iapetus (17$^{\circ}$), Charon (120$^{\circ}$),  and Triton (130$^{\circ}$).

\subsubsection{TTV-TDV diagrams of exoplanets with multiple exomoons in MMR}

\begin{figure*}[!h]
\includegraphics[width=.38\linewidth, angle=-90]{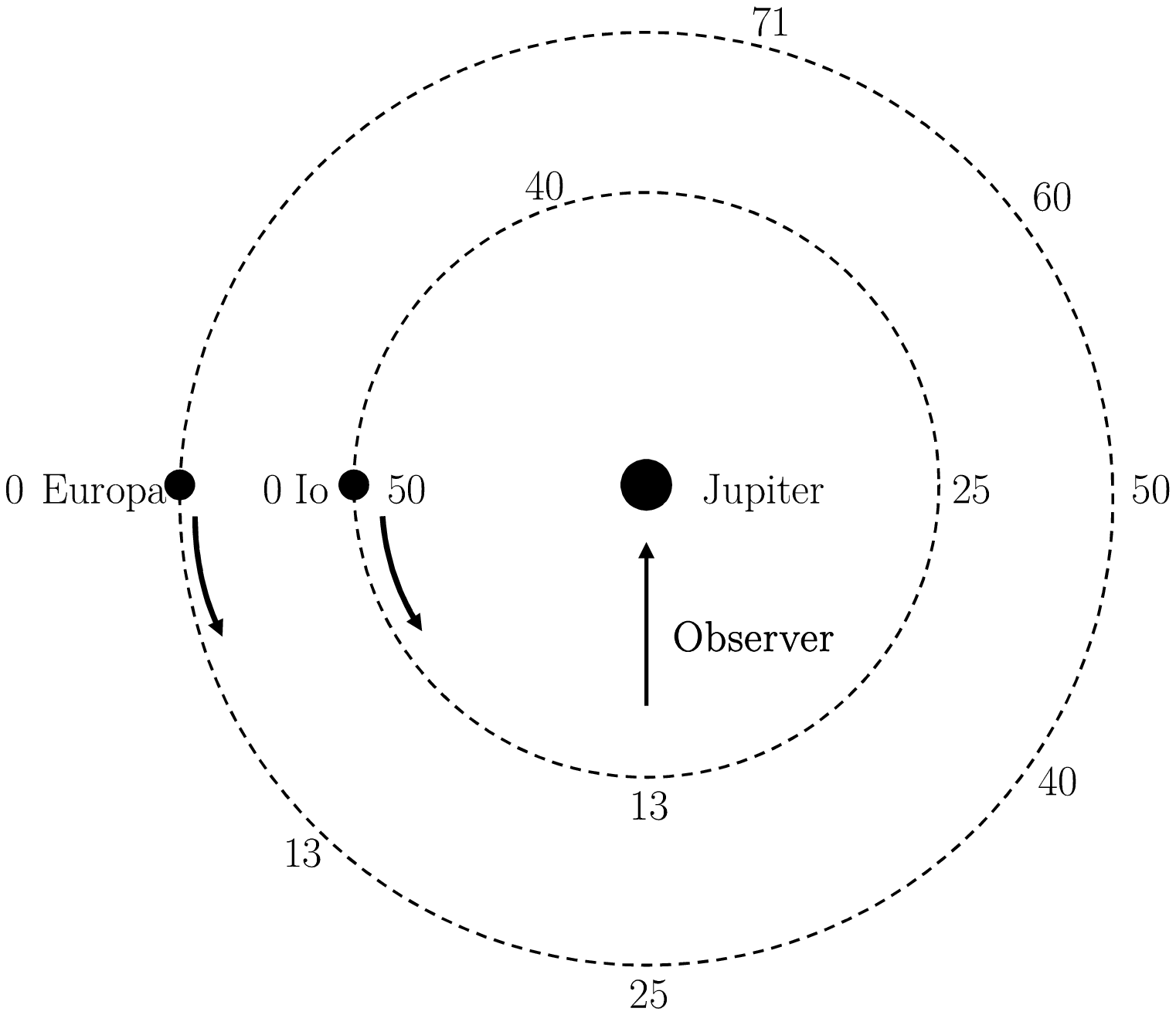}
\hspace{0.25cm}
\includegraphics[width=.41\linewidth, angle=-90]{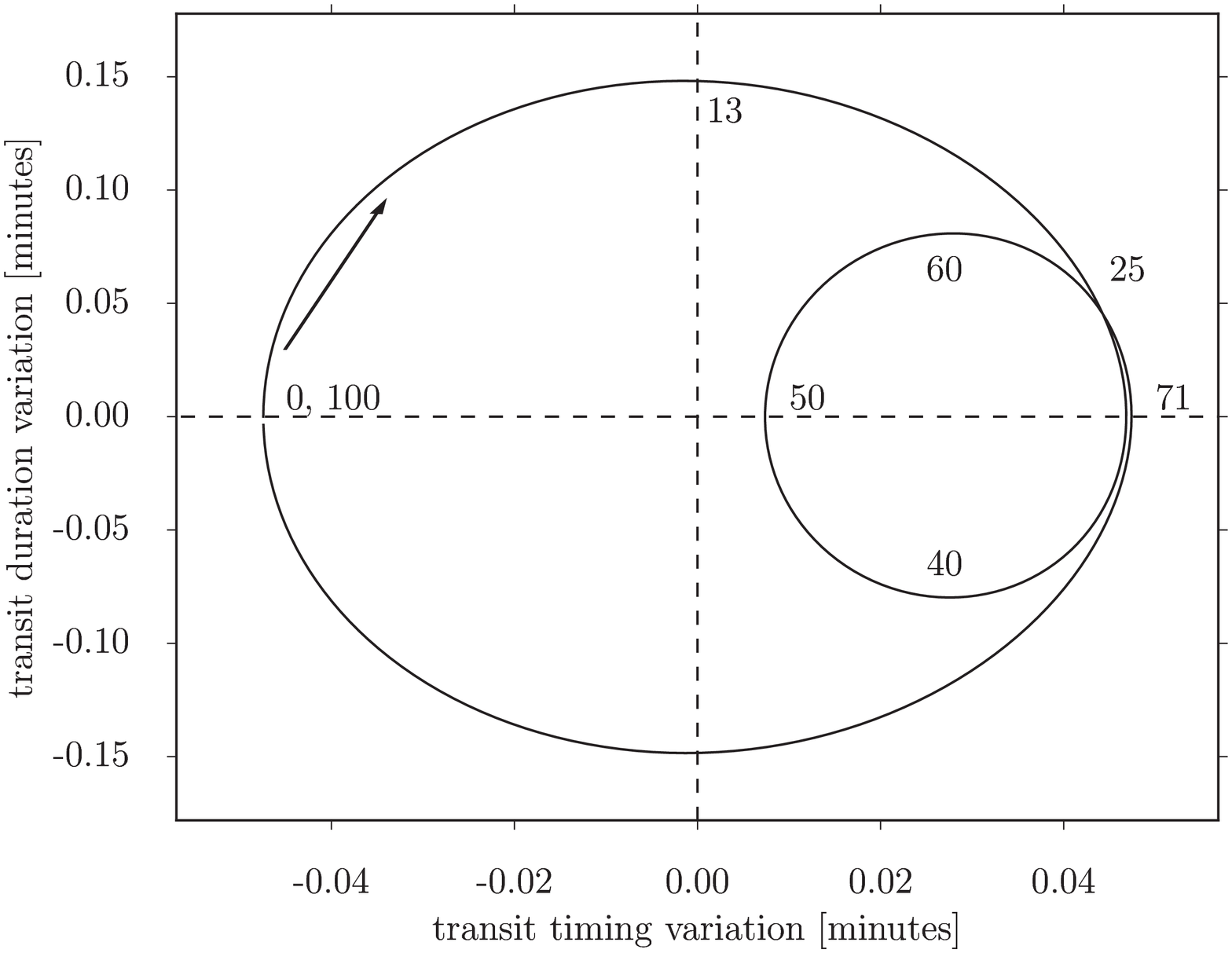}
\caption{\label{fig:howto} Jupiter-Io-Europa system in a 2:1 MMR. Numbers denote the percental orbital phase of the outer moon, Europa. \textit{Left}: Top-down perspective on our two-dimensional $N$-body simulation. The senses of orbital motion of the moons are indicated with curved arrows along their orbits. \textit{Right}: TDV-TTV diagram for the same system. The progression of the numerical TTV-TDV measurements is clockwise.}
\end{figure*}

In Fig.~\ref{fig:howto}, we show the dynamics of a two-moon system around a Jupiter-like planet, where the moons are analogs of Io and Europa. The left panel shows the actual orbital setup in our $N$-body code, the right panel shows the outcome of the planetary TTV-TDV curve over one orbit of the outermost moon in a Europa-wide orbit. Numbers along this track refer to the orbital phase of the outer moon in units of percent, with 100 corresponding to a full orbit. In the example shown, both moons start in a conjunction that is perpendicular and ``to the left'' with respect to the line of sight of the observer (left panel), thereby causing a maximum barycentric displacement of the planet ``to the right'' from the perspective of the observer. Hence, if the planet were to transit the star in this initial configuration of our setup, its transit would occur too early with respect to the average transit period and with a most negative TTV (phase 0 in Fig.~\ref{fig:howto}).

As an analytical check, we calculate the orbital phases of the outer moon at which the planetary motion reverses. The planetary deflection due to the Io-like moon is of functional form $f_{\rm I}(t)~\propto~-k_{\rm I}\cos(n_{\rm I}t)$, where $n_{\rm I}$ is the orbital mean motion of Io, $k_{\rm I}$ is the amplitude of the planetary reflex motion due to the moon, and $t$ is time. The planetary displacement due to the Europa-like moon is $f_{\rm E}(t)~\propto~-k_{\rm E}\cos(n_{\rm E}t)$, where $n_{\rm E}$ is the orbital mean motion of Europa and $k_{\rm E}$ is the amplitude. The total planetary displacement then is $f_{\rm tot}(t)~\propto~f_{\rm I}(t)+f_{\rm E}(t)$ and the extrema can be found, where

\begin{equation}\label{eq:extrema}
0 \stackrel{!}{=} \frac{d}{dt} f_{\rm tot}(t) \ \ ,
\end{equation}

\noindent
that is, where the planetary tangential motion reverses and the planet ``swings back''. We replace $n_{\rm I}$ with $2n_{\rm E}$ and $k_{\rm I}$ with $k_{\rm E}$ (since $k_{\rm I}=1.17\,k_{\rm E}$) and find that Eq.~(\ref{eq:extrema}) is roughly equivalent to

\begin{equation}
0 \stackrel{!}{=} 2\sin(2n_{\rm E}t) + \sin(n_{\rm E}t) \ \ ,
\end{equation}

\noindent
which is true at $n_{\rm E}t~=~0$, $1.824$, $\pi$, $4.459$, $2\pi$, $8.1$ etc., corresponding to orbital phases $n_{\rm E}t/(2\pi)~=~0\,\%, 29\,\%$, $50\,\%$, $71\,\%$, $100\,\%$, $129\,\%$. These values agree with the numerically derived values at the maximum displacements (Fig.~\ref{fig:howto}, right panel) and they are independent of the spatial dimensions of the system.

\begin{figure}[!h]
\centering
\includegraphics[width=.99\linewidth]{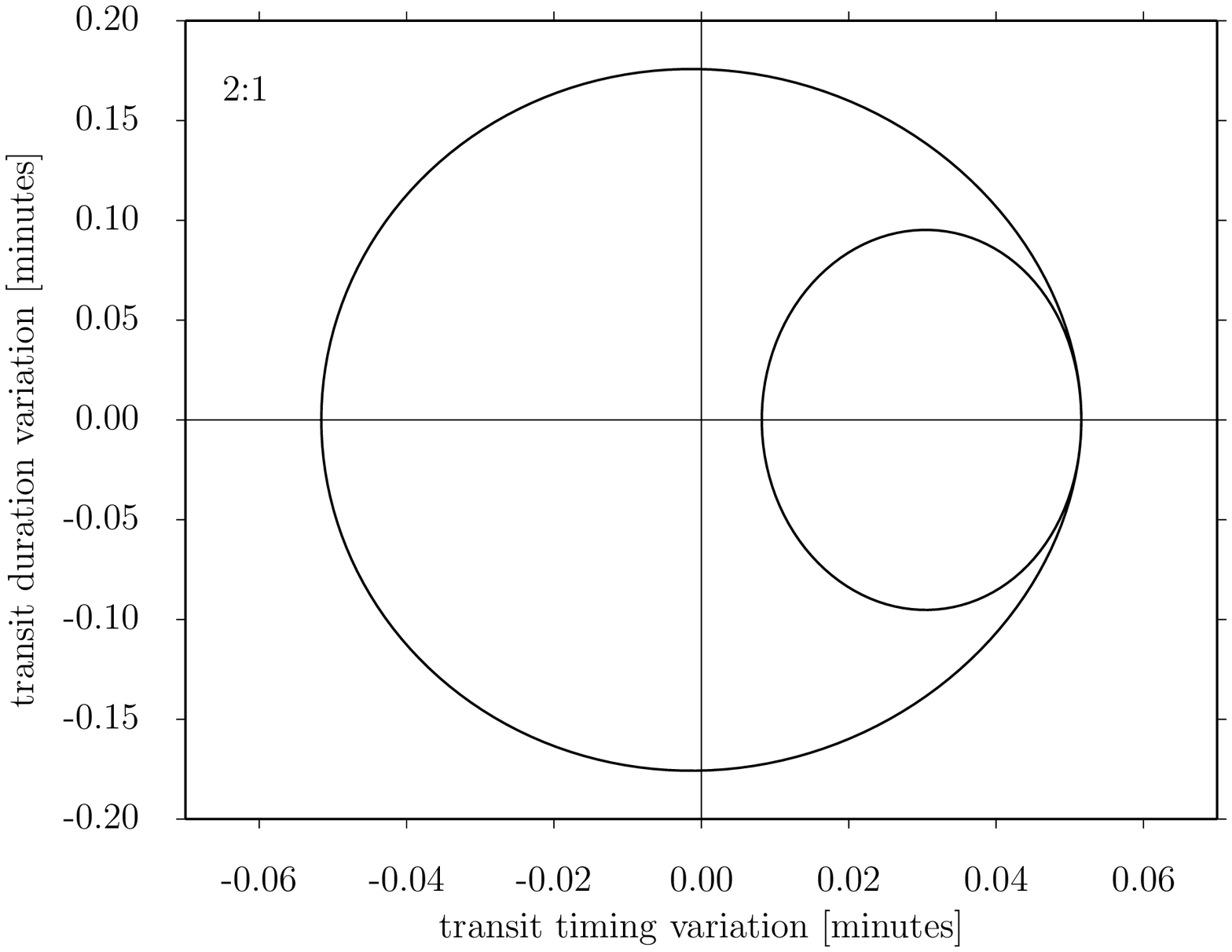}
\includegraphics[width=.99\linewidth]{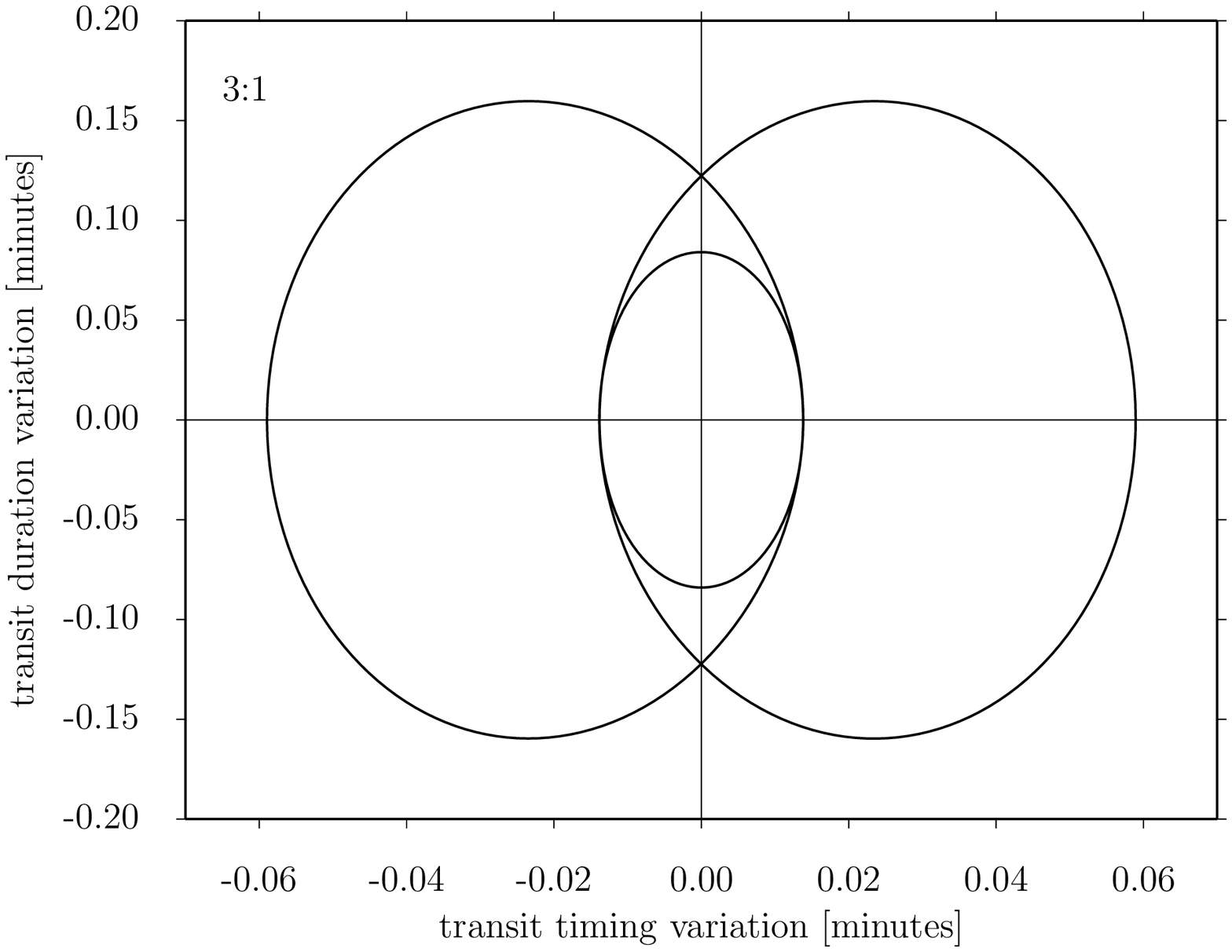}
\includegraphics[width=.99\linewidth]{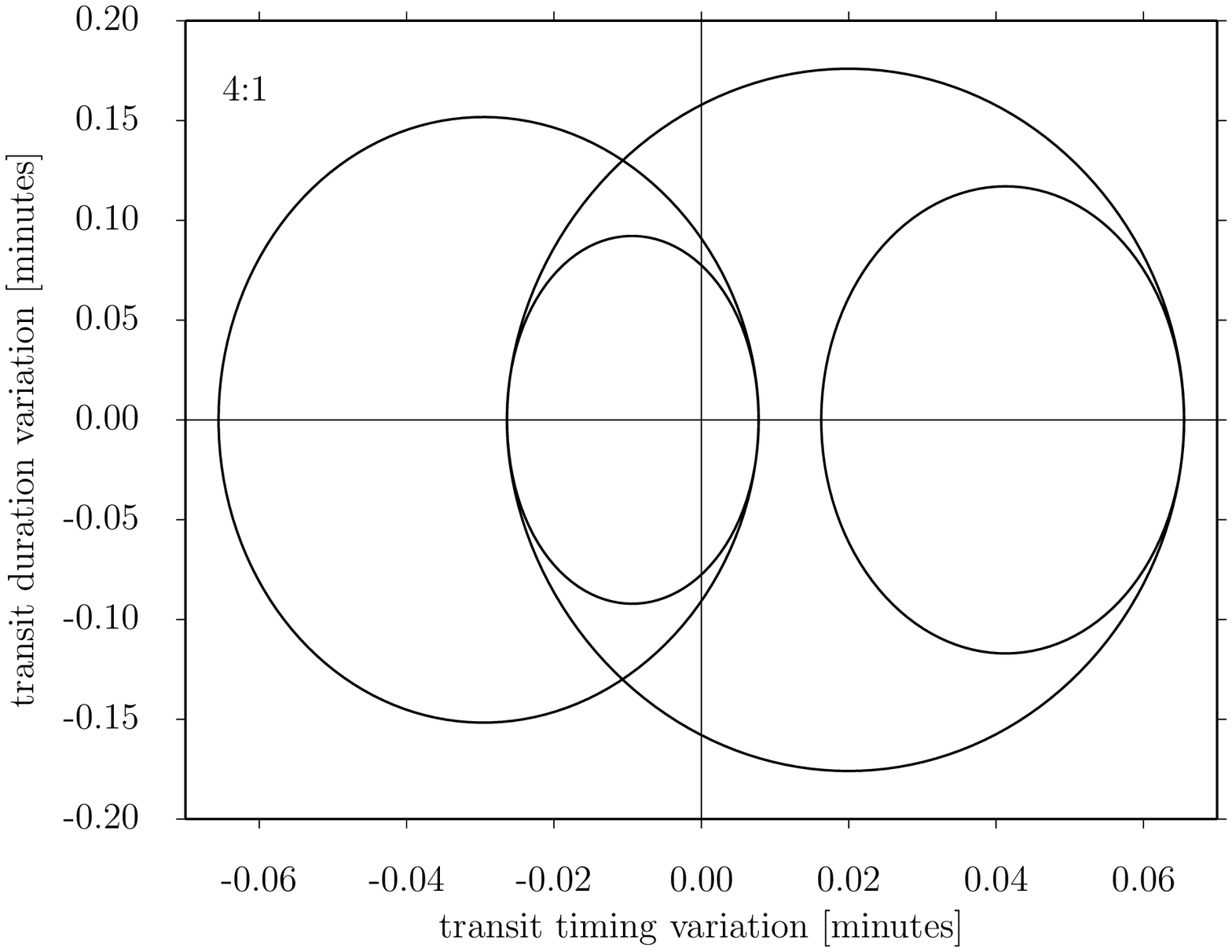}
\caption{\label{fig:period_ratios} TTV-TDV diagrams for a Jupiter-like planet with 2 moons akin to Io and Europa, but in different MMRs. A 2:1 MMR (top) produces 2 ellipses, a 3:1 MMR (center) 3, and a 4:1 MMR (bottom) 4.}
\end{figure}

Next, we explore different orbital period ratios in two-moon MMR systems. In Fig.~\ref{fig:period_ratios}, we show the TTV-TDV diagrams for a Jupiter-like planet (5.2\,au from a Sun-like star) with an Io-like plus a Europa-like moon, but in a 3:1 MMR (top panel) and a 4:1 MMR (bottom panel). Intriguingly, we find that the order of the MMR, or the largest integer in the MMR ratio, determines the number of loops in the diagram. Physically speaking, a loop describes the reverse motion of the planet due to the reversal of one of its moons. This behavior is observed in all our numerical simulations of MMRs of up to five moons (see Appendix \ref{sec:app}).

\begin{figure*}[!h]
\includegraphics[width=.5\linewidth]{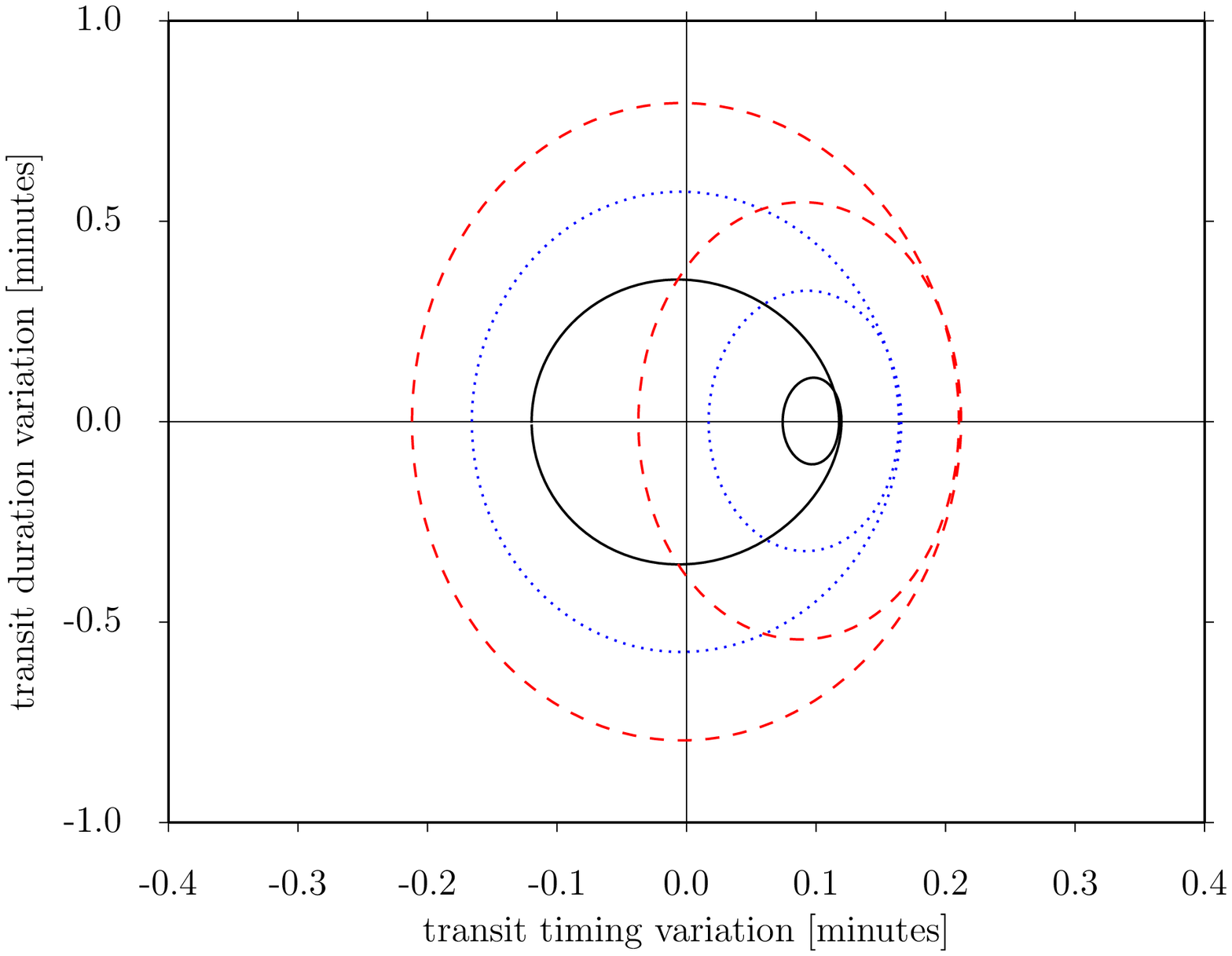}
\includegraphics[width=.5\linewidth]{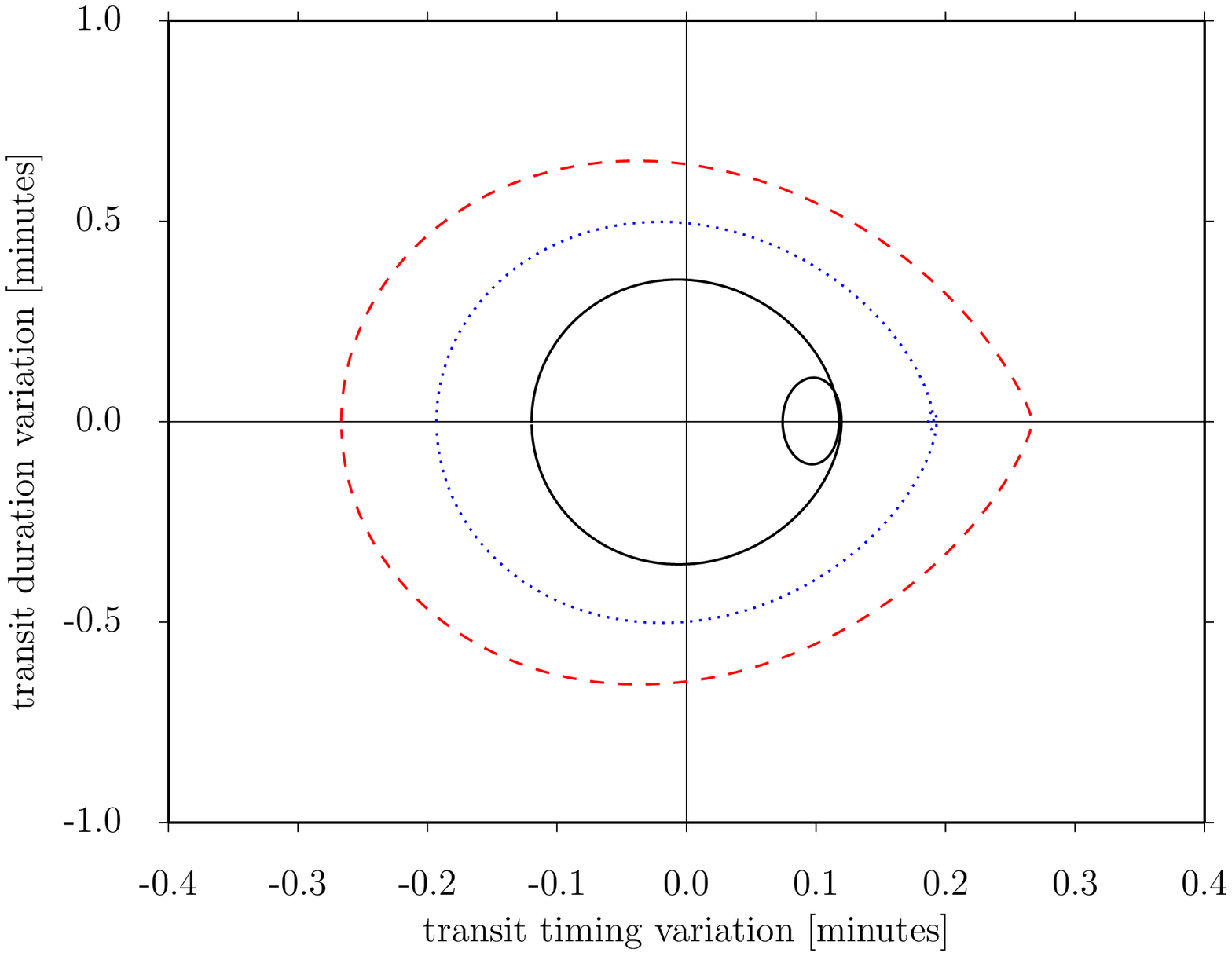}\\\\
\includegraphics[width=.5\linewidth]{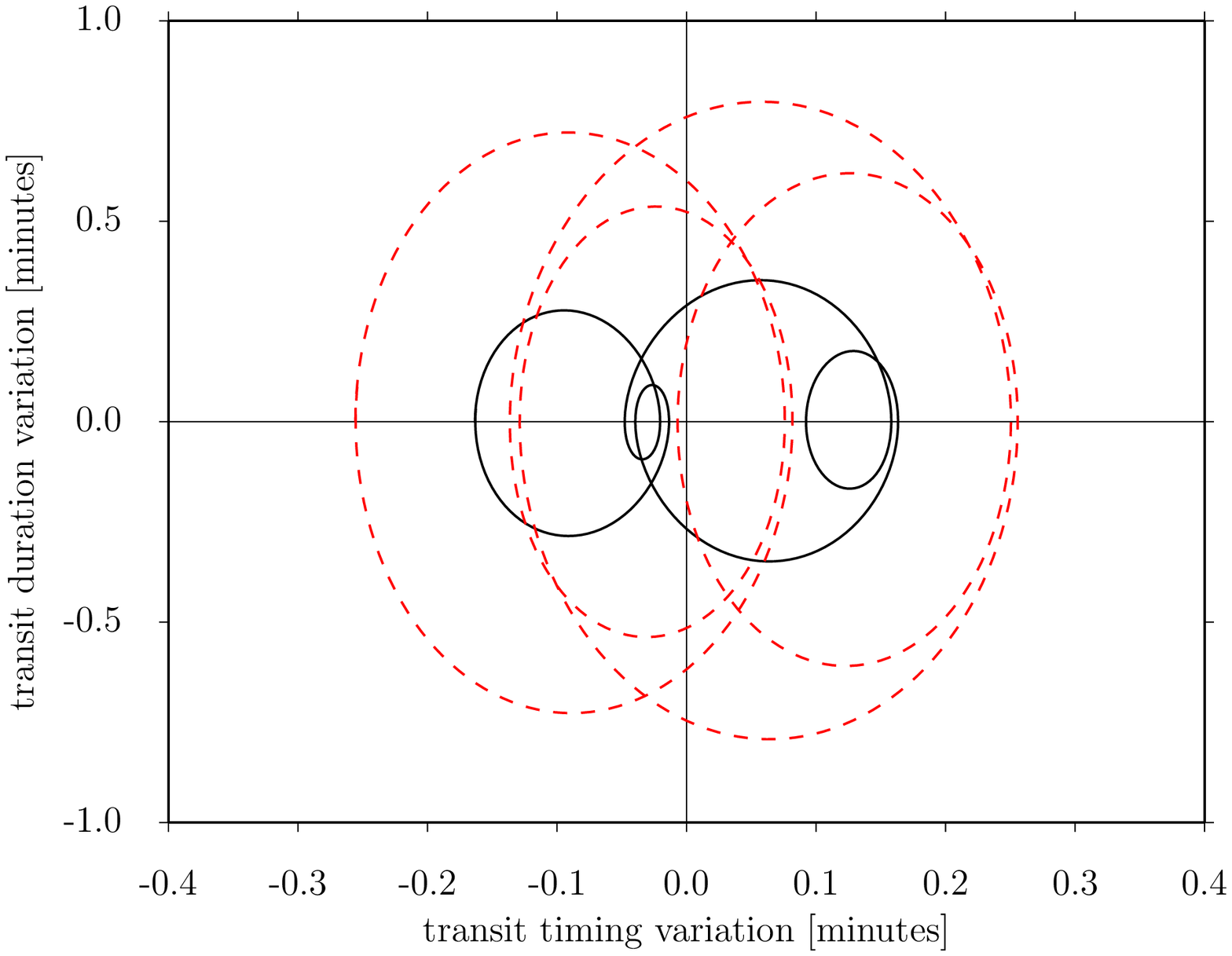}
\includegraphics[width=.5\linewidth]{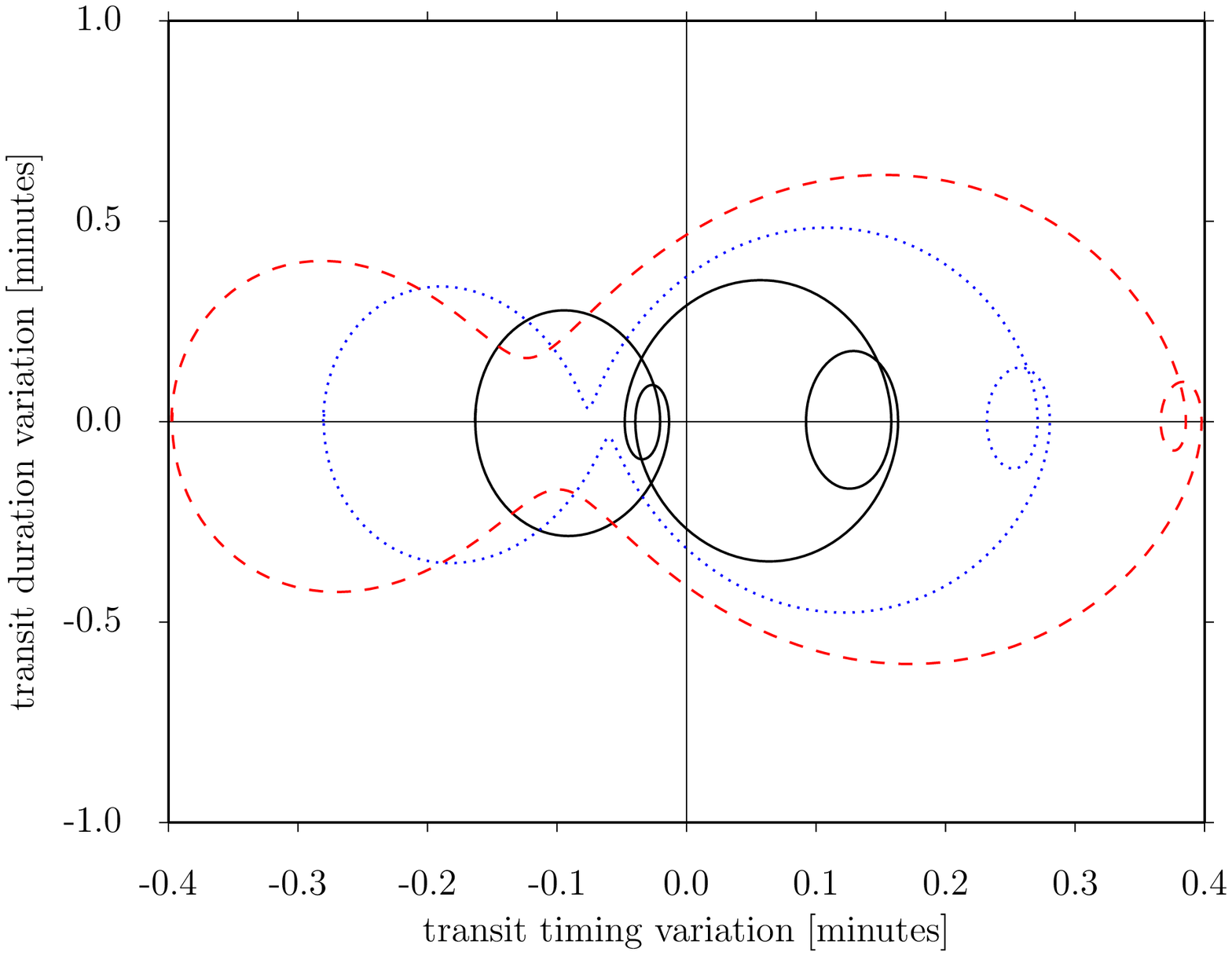}
\caption{\label{fig:masses} TTV-TDV diagrams of a Jupiter-like planet with 2 moons. Top panels assume a 2:1 MMR akin to the Io-Europa resonance; bottom panels assume a 4:1 MMR akin to the Io-Ganymede resonance. Models represented by black solid lines assume that both moons are as massive as Ganymede. In the left panels, blue dotted lines assume the inner moon is twice as massive ($M_{\rm s1}=2\,M_{\rm s2}=2\,M_{\rm Gan}$), and red dashed lines assume the inner moon is thrice as massive. In the right panels, blue assumes the outer moon is twice as massive ($M_{\rm s2}=2\,M_{\rm s1}$) and red assumes the outer moon is thrice as massive. The inner loop becomes invisibly small for low $M_{\rm s1}/M_{\rm s2}$ ratios in the top right panel.}
\end{figure*}

The sizes of any of these loops depends on the moon-to-planet mass ratios and on the semimajor axes of the moons. This dependency is illustrated in Fig.~\ref{fig:masses}, where we varied the masses of the moons. In the upper (lower) panels, a 2:1 (4:1) MMR is assumed. In the left panels, the mass of the outer moon ($M_{\rm s2}$) is fixed at the mass of Ganymede ($M_{\rm Gan}$), while the mass of the inner moon ($M_{\rm s1}$) is successively increased from $1\,M_{\rm Gan}$ (black solid line) over $2\,M_{\rm Gan}$ (blue dotted line) to $3\,M_{\rm Gan}$ (red dashed line). In the right panels, the mass of the inner moon is fixed at $1\,M_{\rm Gan}$, while the mass of the outer moon is varied accordingly. As an important observation of these simulations, we find that massive outer moons can make the loops extremely small and essentially undetectable.

\subsubsection{Evolution of TTV-TDV diagrams in multiple moon systems}
\label{sec:twomoons}

Orbital MMRs can involve librations of the point of conjunction as well as drifts of the pericenters of the moons. In fact, the perijoves (closest approaches to Jupiter) in the Io-Europa 2:1 MMR shows a drift of about $0^\circ.7\,{\rm day}^{-1}$. Hence, the MMR is only valid in a coordinate system that rotates with a rate equal to the drift of the pericenters of the respective system. The libration amplitude of the pericenters on top of this drift has been determined observationally to be $0^\circ.0247\pm0^\circ.0075$ \citep{1928AnLei..16B...1D}. For multiple exomoon systems, these effects would smear the TTV-TDV figures obtained with our simulations, if the drift is significant on the timescale on which the measurements are taken.

\begin{figure}[!h]
\includegraphics[width=1.\linewidth]{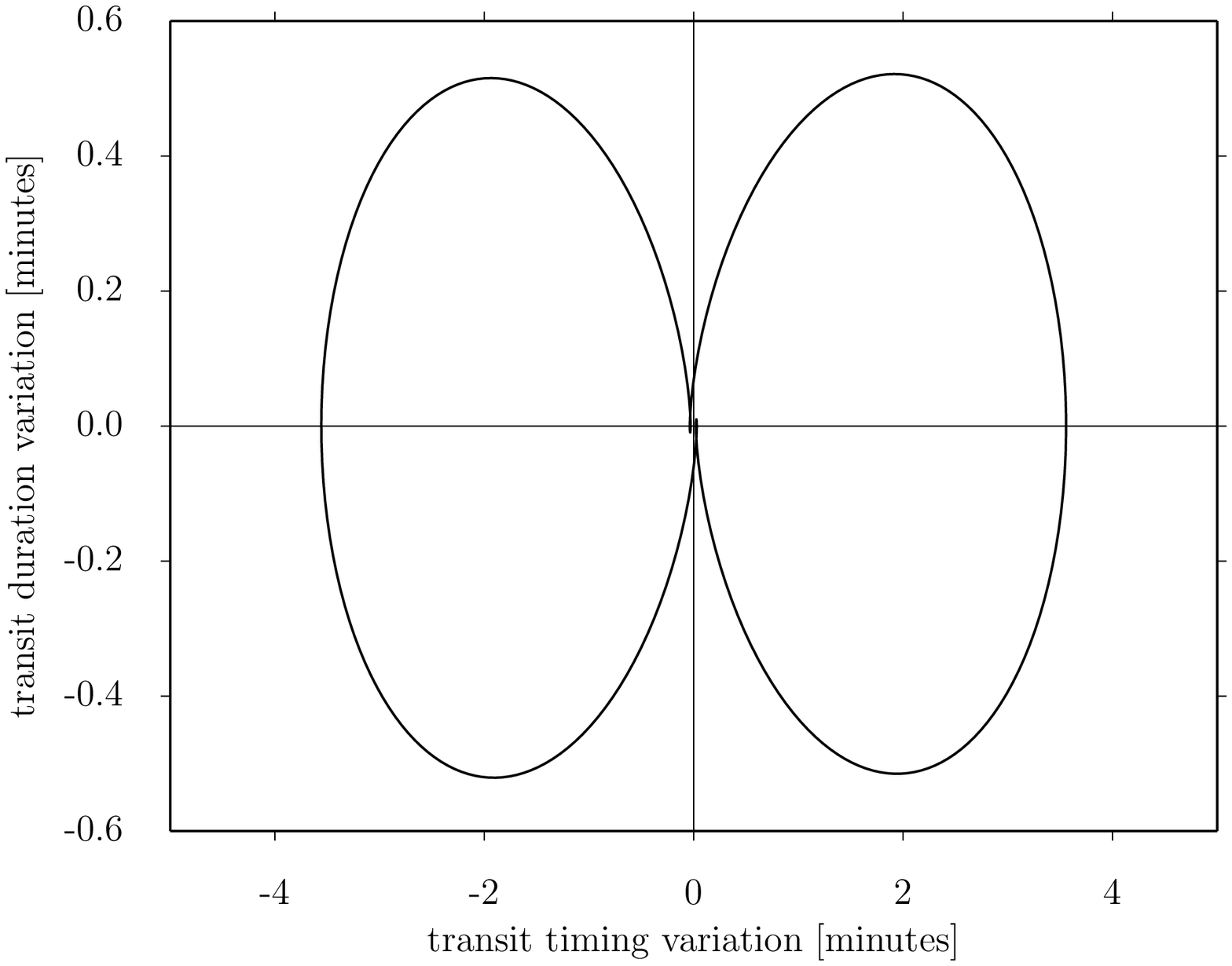}
\includegraphics[width=1.\linewidth]{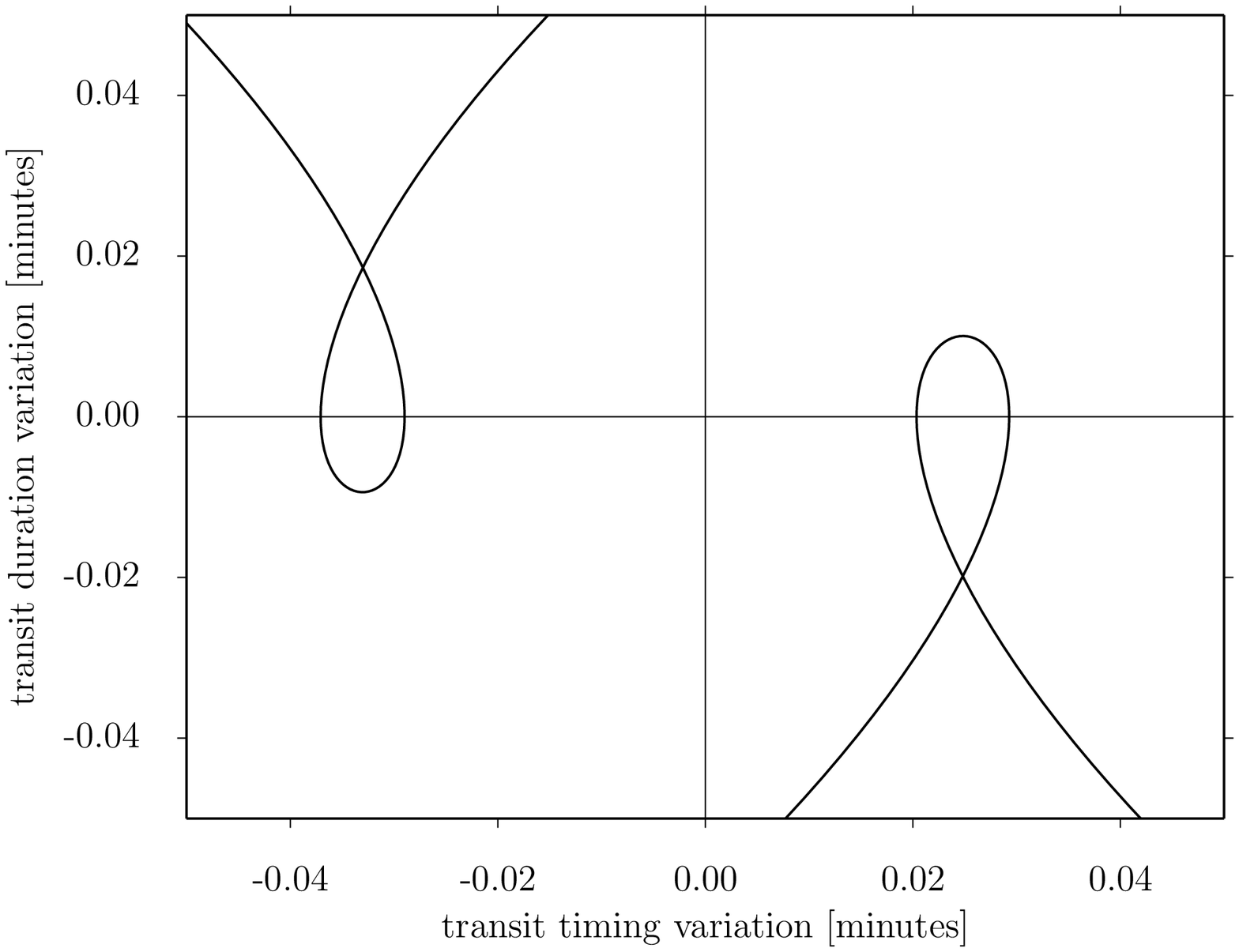}
\includegraphics[width=1.\linewidth]{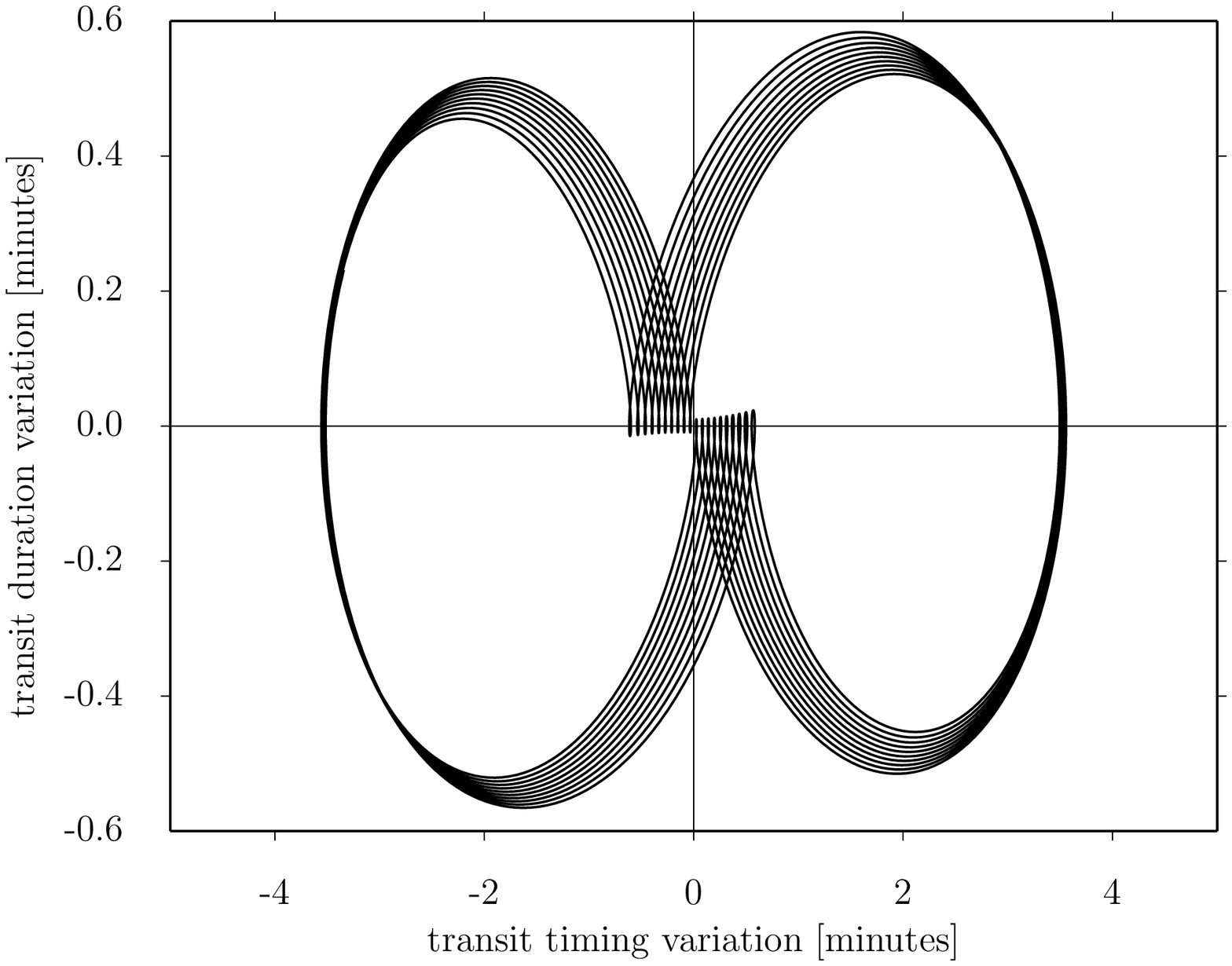}
\caption{\label{fig:twomoons} \textit{Top}: TTV-TDV diagram the Earth-Moon-like system involving a second satellite at half the distance of the Moon to Earth with 70\,\% of the lunar mass. \textit{Center}: Zoom into the center region showing the turning points. \textit{Bottom}: After sampling 10 orbits, a smearing effect becomes visible as the 2 moons are not in a MMR. Exomoon retrieval from such a more complex TTV-TDV figure needs to take into account the evolution of the system.}
\end{figure}

If the moons are not in a MMR in the first place, then there is an additional smearing effect of the TTV-TDV figures because of the different loci and velocities of all bodies after one revolution of the outermost moon. We explore this effect using an arbitrary example, in which we add a second moon to the Earth-Moon system. We choose an arbitrary semimajor axis (50\,\% lunar) and mass (0.475\% lunar), and reduced the Moon's mass to 70\,\% lunar. The upper panel in Fig.~\ref{fig:twomoons} shows the TTV-TDV diagram after a single orbit of the outermost moon. As this system does not include a low-integer MMR, the TTV-TDV figure is very complex, involving an hourglass shape main figure with two minute loops around the origin (Fig.~\ref{fig:twomoons}, center panel). As expected, the resulting TTV-TDV diagram after multiple moon orbits exhibits a smearing effect (Fig.~\ref{fig:twomoons}, bottom panel). In this particular example system, the overall shape of the TTV-TDV figure actually remains intact, but much more substantial smearing may occur in other systems.

\section{Blind retrieval of single and multiple moon systems}
\label{sec:blind}

Next, we want to know whether the above-mentioned TTV-TDV patterns can actually be detected in a realistic dataset. Above all, observations only deliver a limited amount of TTV-TDV measurements per candidate system and white noise and read noise introduce uncertainties. To which extent does real TTV-TDV data enable the detection of exomoons, and permit us to discern single from multiple exomoon systems?

A $\chi^2$ test can determine the best-fitting model if the model parameters are known. In a realistic dataset, however, the number of parameters is generally unknown since the number of moons is unknown. Hence, we performed a Bayesian test, in which two of us (MH, RH) prepared datasets that were then passed to the other coauthors (BP, DA) for analyses. The preparation team kept the number of moons in the data secret but constrained it to be either 0, 1, or 2. It was agreed that any moons would be in circular, prograde, and stable orbits, that is, beyond the Roche radius but within $0.5\,R_{\rm H}$ \citep{2006MNRAS.373.1227D}. Our code allow us to simulate and retrieve eccentric moons as well, but for demonstration purpose we restrict ourselves to circular orbits in this study. Yet, eccentric and inclined orbits would need to be considered in a dedicated exomoon survey.

\subsection{Parameterization of the planetary system}
\label{subsec:system}

We chose to use an example loosely based on KOI-868, a system searched for exomoons by \citet{2015ApJ...813...14K}. We kept $P_{\rm p}=236$\,d, $M_\star=0.55\pm0.07\,M_{\odot}$, and $R_\star=0.53\pm0.07\,R_\odot$, and the measured timing errors of $\sim1.5$\,min per data point. A preliminary analysis suggested that a Moon-like moon could not possibly be detected about KOI-868\,b with our approach as the TTV-TDV signals of the $0.32\,M_{\rm Jup}$ planet would be too small. Hence, we assumed an Earth-mass planet ($M_{\rm p}=1\,M_{\oplus}\pm0.34$). The mass uncertainty of the planet is based on realistic uncertainties from Earth-sized planets, for example, Kepler-20\,f \citep{2012ApJ...749...15G} and requires radial velocity measurements of better than 1m s$^{-1}$ \citep{2012Natur.482..195F}. Our hypothesized planet would have a much smaller radius than KOI-868\,b, potentially resulting in less accurate timing measurements than we assumed. This neglect of an additional source of noise is still reasonable, since there is a range of small \textit{Kepler} planets with very precise timing measurements, for example, Kepler-80\,d with a radius of $R_{\rm p}=1.7\pm0.2\,R_{\oplus}$ and timing uncertainties $\sim1$\,min \citep{2014ApJ...784...45R}. To produce consistent TTV-TDV simulations, we adjust the transit duration to $0.3$\,d.

\subsection{Parameterization of the one-moon system}

In our first example, we assumed a heavy moon ($0.1\,M_{\oplus}$) in a stable, circular Moon-wide orbit ($a_{\rm s}=3.84\times10^5\,{\rm km}$ or 34\,\% the Hill radius of this planet, $R_{\rm H}$). This high a mass yields a moon-to-planet mass ratio that is still slightly smaller than that of the Pluto-Charon system. Its sidereal period is 26.2\,d, slightly shorter than the 27.3\,d period of the Moon. The resulting TTV and TDV amplitudes are $23.1$\,min and $\pm1.6$\,min, respectively. A total of seven data points were simulated, which is consistent with the number of measurements accessible in the four years of \textit{Kepler} primary observations. The data were simulated using our $N$-body integrator, then stroboscopically spread over the TTV-TDV diagram (see Sect.~\ref{sec:single}) and randomly moved in the TTV-TDV plane assuming Gaussian noise. The retrieval team treated $M_\star$, $R_\star$, $M_{\rm p}$, and $a_{\rm B}$ as fixed and only propagated the errors of $R_\star$ and $M_{\rm p}$, which is reasonable for a system that has been characterized spectroscopically.

\subsection{Parameterization of the two-moon system}

Our preliminary simulations showed that multiple exomoon retrieval based on \textit{Kepler}-style data quality only works in extreme cases with moons much larger than those known from the solar system or predicted by moon formation theories. Hence, we assumed an up-scaled space telescope with a theoretical instrument achieving ten times the photon count rate of \textit{Kepler}, corresponding to a mirror diameter of $\approx3.3$\,m (other things being equal); this is larger than the \textit{Hubble Space Telescope} (2.5\,m), but smaller than the \textit{James Webb Space telescope} (6.5\,m). Neglecting other noise sources such as stellar jitter, our hypothetical telescope reaches ten times higher cadence than \textit{Kepler} at the same noise level. We also found that seven data points do not sample the TTV-TDV figure of the planet sufficiently to reveal the second moon. Rather more than thrice this amount is necessary. We thus simulated 25 data points, corresponding to 15 years of observations. This setup is beyond the technological capacities that will be available within the next decade or so, and our investigations of two-moon systems are meant to yield insights into the principal methodology of multiple moon retrieval using TTV-TDV diagrams.

Our $N$-body simulations suggested that the second moon cannot occupy a stable, inner orbit if the more massive, outer moon has a mass $\gtrsim0.01\,M_{\rm p}$, partly owing to the fact that both moons must orbit within $0.5\,R_{\rm H}$. We neglect exotic stable configurations such as the Klemperer rosette \citep{1962AJ.....67..162K} as they are very sensitive to perturbations. Instead, we chose masses of $0.02\,M_{\oplus}$ for the inner moon and $0.01\,M_{\oplus}$ for the outer moon, which is about the mass of the Moon. We set the semimajor axis of the outer moon ($a_{\rm s2}$) to $1.92\times10^5\,{\rm km}$, half the value of the Earth's moon. The inner moon was placed in a 1:2 MMR with a semimajor axis ($a_{\rm s1}$) of $1.2\times10^5\,{\rm km}$. The masses of the inner and outer satellites are referred to as $M_{\rm s1}$ and $M_{\rm s2}$, respectively.

\subsection{Bayesian model selection and likelihood}

In order to robustly select the model that best describes the given data, we employ Bayesian model selection \citep{sivia2006data, Knuth201550}, which relies on the ability to compute the Bayesian evidence

\begin{equation}
Z = \int_\theta \pi(\theta) \times L(\theta)  d\theta.
\end{equation}

\noindent
Here, $\pi(\theta)$ represents the prior probabilities for model parameters $\theta$ and quantifies any knowledge of a system prior to analyzing data. $L(\theta)$ represents the likelihood function, which depends on the sum of the square differences between the recorded data and the forward model.
 
The evidence is an ideal measure of comparison between competing models as it is a marginalization over all model parameters. As such, it naturally weighs the favorability of a model to describe the data against the volume of the parameter space of that model and thus aims to avoid the overfitting of data. It can be shown that the ratio of the posterior probabilities for two competing models with equal prior probabilities is equal to the ratio of the Bayesian evidence for each model. Therefore the model with the largest evidence value is considered to be more favorable to explain the data \citep{Knuth201550}.
 
We utilize the MultiNest algorithm \citep{2008MNRAS.384..449F, 2009MNRAS.398.1601F, 2011MNRAS.415.3462F, 2013arXiv1306.2144F} 
to compute log-evidences used in the model selection process, and posterior samples used for obtaining summary statistics for all model parameters. MultiNest is a variant on the Nested Sampling algorithm \citep{skilling2006} and is efficient for sampling within many dimensional spaces that may or may not contain degeneracies. Nested sampling algorithms are becoming increasingly useful in exoplanet science, e.g. for the analysis of transit photometry \citep{2012ApJ...750..115K,2014ApJ...795..112P,2015ApJ...814..147P} or for the retrieval of exoplanetary atmospheres from transit spectroscopy \citep{2013ApJ...778..153B,2015ApJ...802..107W}.

\begin{table}[t!]
\center
\caption{Normalized Bayesian evidence for our simulated datasets as per Eq.~(\ref{eq:fitting}).\label{tab:modelprob}}
\begin{tabular}{ccc}
\hline
no. of moons in model & 1-moon dataset & 2-moon dataset \\
\hline
0 & 0\,\%    &  0\,\%       \\
1 & 57.2\,\% &  0.00004\,\% \\
2 & 42.8\,\% & 99.99996\,\% \\
\hline
\end{tabular}
\end{table}

\begin{table}[t!]
\center
\caption{Specification and blind-fitting results for the one-moon system.\label{tab:singlemoon}}
\begin{tabular}{lcccc}
\hline
Parameter & True & Fitted & \\
\hline
$M_{\rm s1}$  & $0.1\,M_\oplus$ & $0.12 \pm0.03\,M_\oplus$\\
$a_{\rm s1}$ & $384,399$\,km & $413,600\pm 88,450$\,km\\
\hline
\end{tabular}
\end{table}

\begin{table}[t!]
\center
\caption{Specification and blind-fitting results for the two-moon system.\label{tab:multi-moon}}
\begin{tabular}{lcccc}
\hline
Parameter & True & Fitted & \\
\hline
$M_{\rm s1}$  & $0.02\,M_\oplus$ & $0.021\,M_\oplus\pm0.002\,M_\oplus$ \\
$M_{\rm s2}$  & $0.01\,M_\oplus$ & $0.0097 \,M_\oplus\pm 0.0012 \,M_\oplus$ \\
$a_{\rm s1}$ & $120,000$\,km & $131,000\pm 16,560$\,km\\
$a_{\rm s2}$ & $192,000$\,km & $183,000\pm11,000$\,km\\
\hline
\end{tabular}
\end{table}

As inputs, MultiNest requires prior probabilities for all model parameters as well as the log-likelihood function. Priors for all model parameters were chosen to be uniform between reasonable ranges. For the single moon case, we explored a range of possible moon masses and semimajor axes with $0~{\geq}~M_{\rm s}~{\geq}~0.2\,M_\oplus$ and $10^4\,{\rm km}~{\geq}~a_{\rm s}~{\geq}~5.5\times10^5$\,km. The lower and upper limits for $a_{\rm s}$ correspond to the Roche lobe and to $0.5\,R_{\rm H}$, respectively. For the two-moon scenario, the prior probabilities for the orbital distances were kept the same but the moon masses were taken to range within $0~{\geq}~M_{\rm s}~{\geq}~0.03\,M_\oplus$. The upper mass limit has been determined by an $N$-body stability analysis.

Since both TDV and TTV signals must be fit simultaneously, a nearest neighbor approach was adopted for the log-likelihood function. For each data point, the nearest neighbor model point was selected, and the log-likelihood computed for that pair. This approach neglects the temporal information contained in the TTV-TDV measurements or, in our case, simulations. A fit of the data in the TTV-TDV plane is similar to a phase-folding technique as frequently used in radial velocity or transit searches for exoplanets. A fully comprehensive data fit would test all the possible numbers of moon orbits during each circumstellar orbit ($n$, see Sect.~\ref{sec:single}), which would dramatically increase the CPU demands. Assuming Gaussian noise for both TTV and TDV signals, the form of the log-likelihood was taken to be

\begin{align}
\log L & = -\frac{1}{2\sigma^2_{\text{TTV}}}\sum_{i=1}^{N}\left(\mathcal{M}_{\text{TTV},i} - \mathcal{D}_{\text{TTV},i}\right)^2 \nonumber\\
 & - \frac{1}{2\sigma^2_{\text{TDV}}}\sum_{i=1}^{N}\left(\mathcal{M}_{\text{TDV},i} - \mathcal{D}_{\text{TDV},i}\right)^2
\end{align}

\noindent
where $\mathcal{M}_{\text{TTV},i}$ and $\mathcal{M}_{\text{TDV},i}$ are the TTV and TDV coordinates of the nearest model points to the $i^{\rm th}$ data points, $\mathcal{D}_{\text{TTV},i}$ and $ \mathcal{D}_{\text{TDV},i}$, $N$ is the number of data points, and $\sigma^2$ is the signal variance. Ultimately, we normalize the evidence for the $i^{\rm th}$ model as per

\begin{equation}\label{eq:fitting}
(\text{normalized evidence})_i = \frac{Z_i}{Z_1 + Z_2 + Z_3}
\end{equation}

\noindent
to estimate the probability that a model correctly describes the data. The left-hand side of Eq.~(\ref{eq:fitting}) would change if we were to investigate models with more than two moons.

\section{Results}
\label{sec:results}

\subsection{Blind retrieval of multiple exomoons}

\begin{figure}[!h]
\centering
\includegraphics[width=0.9\linewidth]{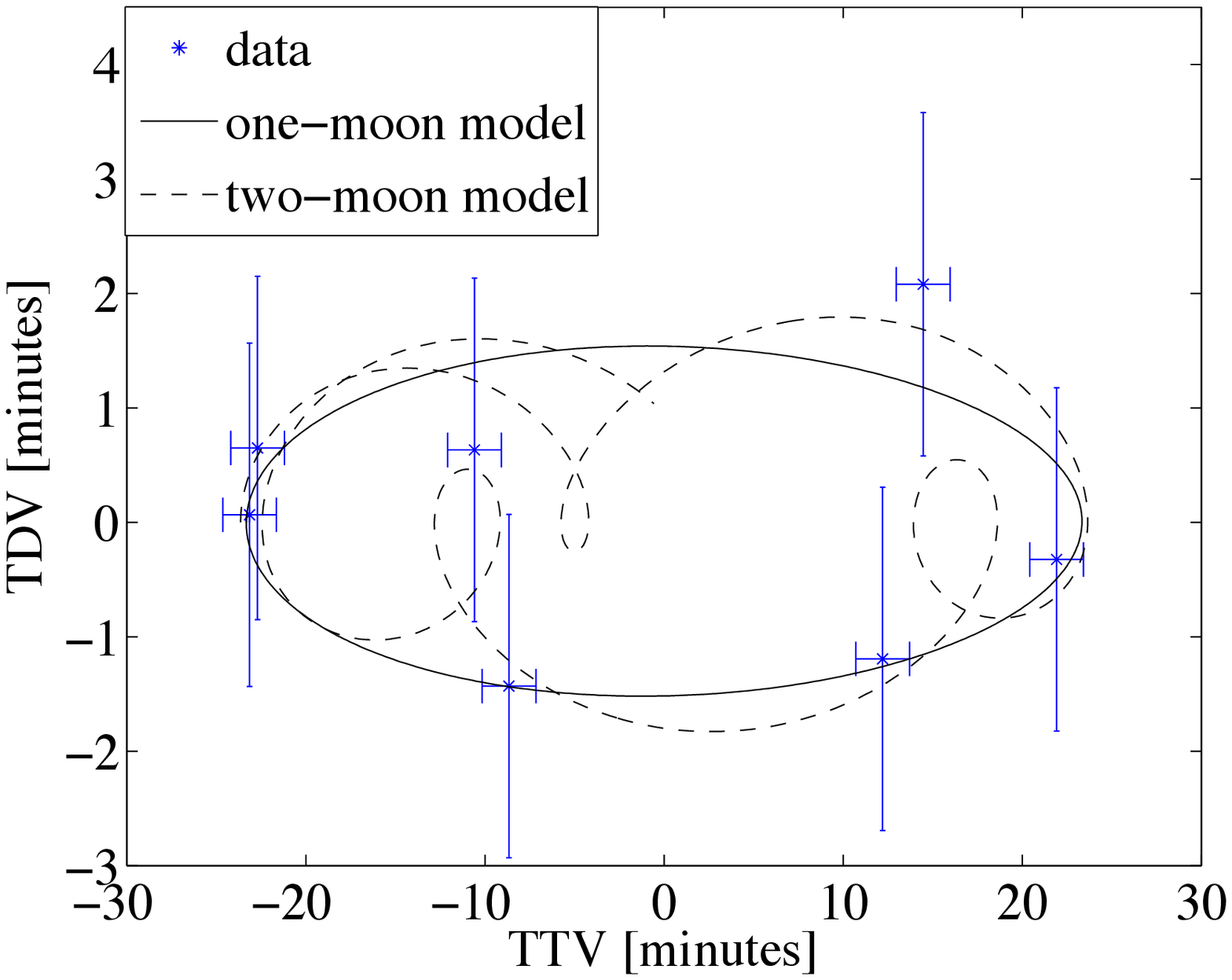}
\caption{\label{fig:singlemoonfits} One- and two-moon fits to the model generated data of one moon.}
\end{figure}

The results of our log-evidence calculations for the blind exomoon retrieval are shwon in Table~\ref{tab:modelprob}. In the case where a one-moon system had been prepared for retrieval, the Bayesian log-evidences are $\log Z_0 = -463.44 \pm 0.02$, $\log Z_1 = -2.98 \pm 0.10$, and $\log Z_2 = -3.27 \pm 0.15$, indicating a slight preference of the one-moon model over both the two-moon and the zero-moon cases. The best fits to the data are shown in Fig.~\ref{fig:singlemoonfits}. While the difference in log-evidence between the one- and two-moon models is small, our retrieval shows that an interpretation with moon, be it a one- or a multiple system, is strongly favored over the planet-only hypothesis. A moon with zero mass is excluded at high confidence.

\begin{figure}[!h]
\centering
\includegraphics[width=0.93\linewidth]{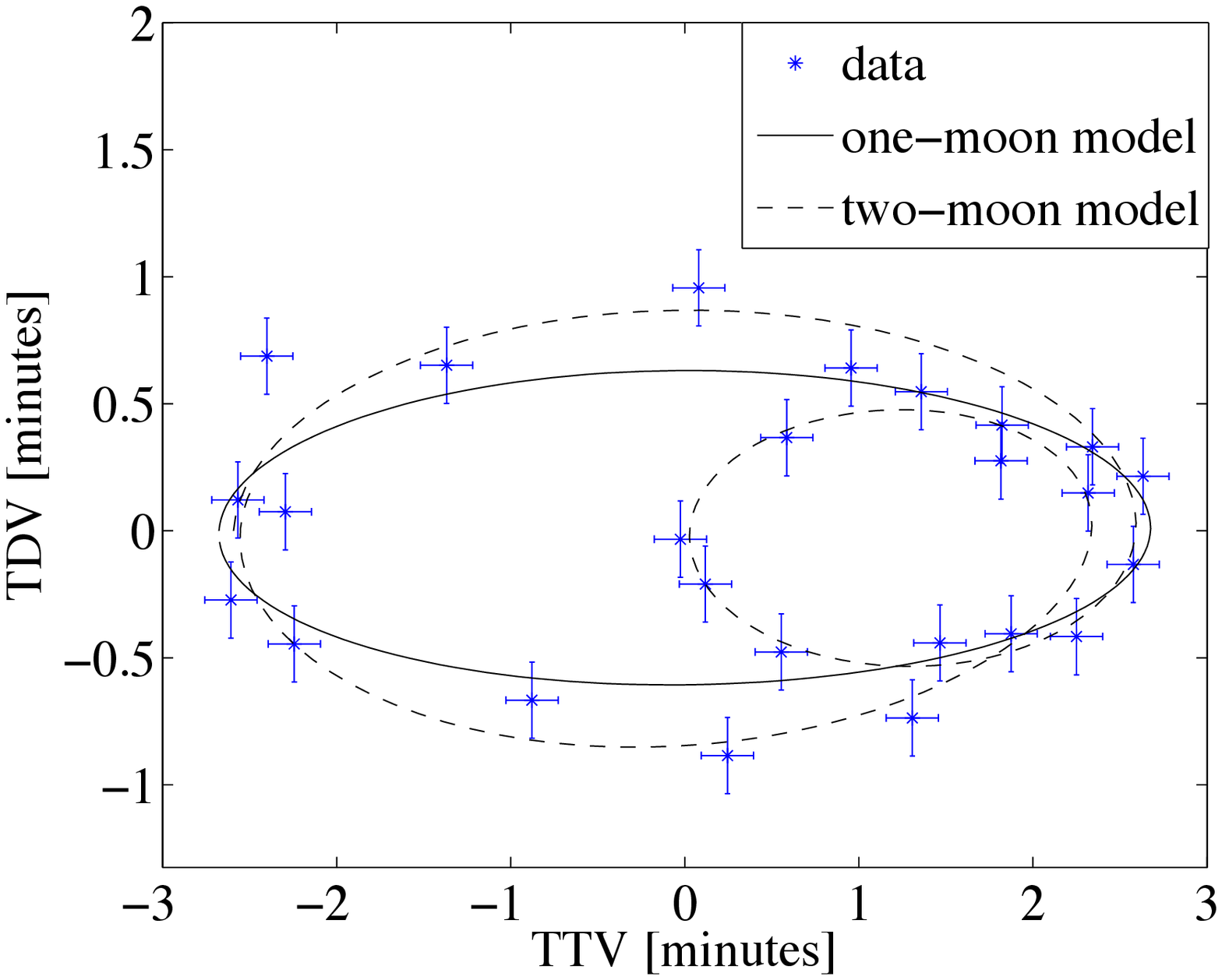}
\caption{\label{fig:multi-moonfits} One- and two-moon fits to the model generated data of two moons. }
\end{figure}

In the two-moon case, the log-evidences for the zero-, one-, and two-moon models are $\log Z_0 = -1894.49 \pm 0.02$, $\log Z_1 = -43.96 \pm 0.37$, and $\log Z_2 = -29.29 \pm 0.47$, respectively. The fits to the data are shwon in Fig.~\ref{fig:multi-moonfits}. In this case, the two-moon model is highly favored over both the zero- and one-moon models. 

\begin{figure}[!h]
\centering
\includegraphics[width = 1.0\linewidth]{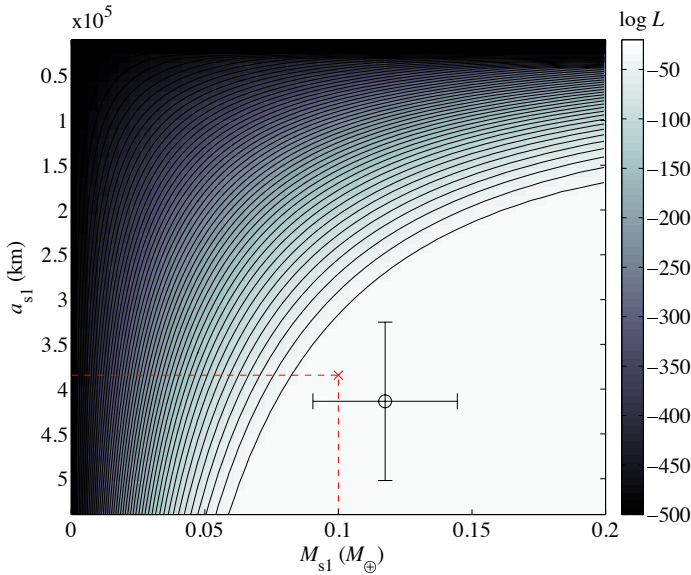}
\caption{\label{fig:probplotssingle} Log-likelihood landscape of the one-moon model applied to the simulated one-moon data.  Lighter areas indicate regions of high probability and the red cross indicates the true parameter values. One can clearly see a degeneracy between $a_{\rm s1}$ and $M_{\rm s1}$ in the form of a curved plateau in the bottom right portion of the plot.}
\end{figure}

Once the most likely number of moons in the system has been determined, we were interested in the parameter estimates for the moons. For the one-moon case, our estimates are listed in Table~\ref{tab:singlemoon}. The one-moon model, which has the highest log-evidence, predicts $M_{\rm s1} = 0.12 \pm 0.03\,M_\oplus$ and $a_{\rm s1} = 413,600 \pm 88,450$\,km. The relatively large uncertainties indicate an $M_{\rm 1s}$--$a_{\rm s1}$ degeneracy. Figure~\ref{fig:probplotssingle} shows the log-likelihood contours of the one-moon model applied to this dataset. Indeed, the curved probability plateau in the lower right corner of the plot suggests a degeneracy between the moon mass and orbital distance, which may be exacerbated by the small amplitude of the TDV signal.

\begin{figure*}[!h]
\centering
\includegraphics[width=.479\linewidth]{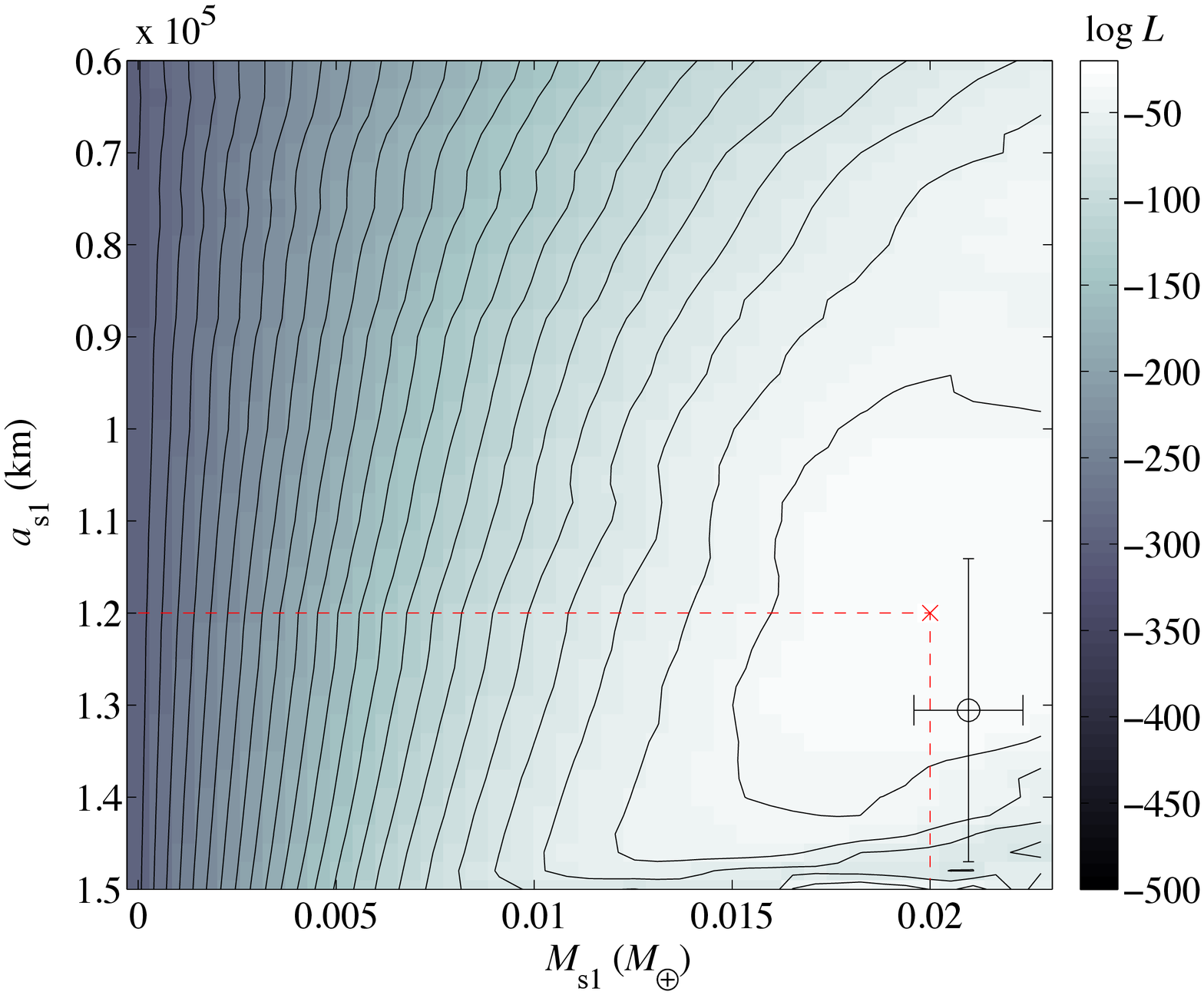}
\hspace{0.229cm}
\includegraphics[width=.488\linewidth]{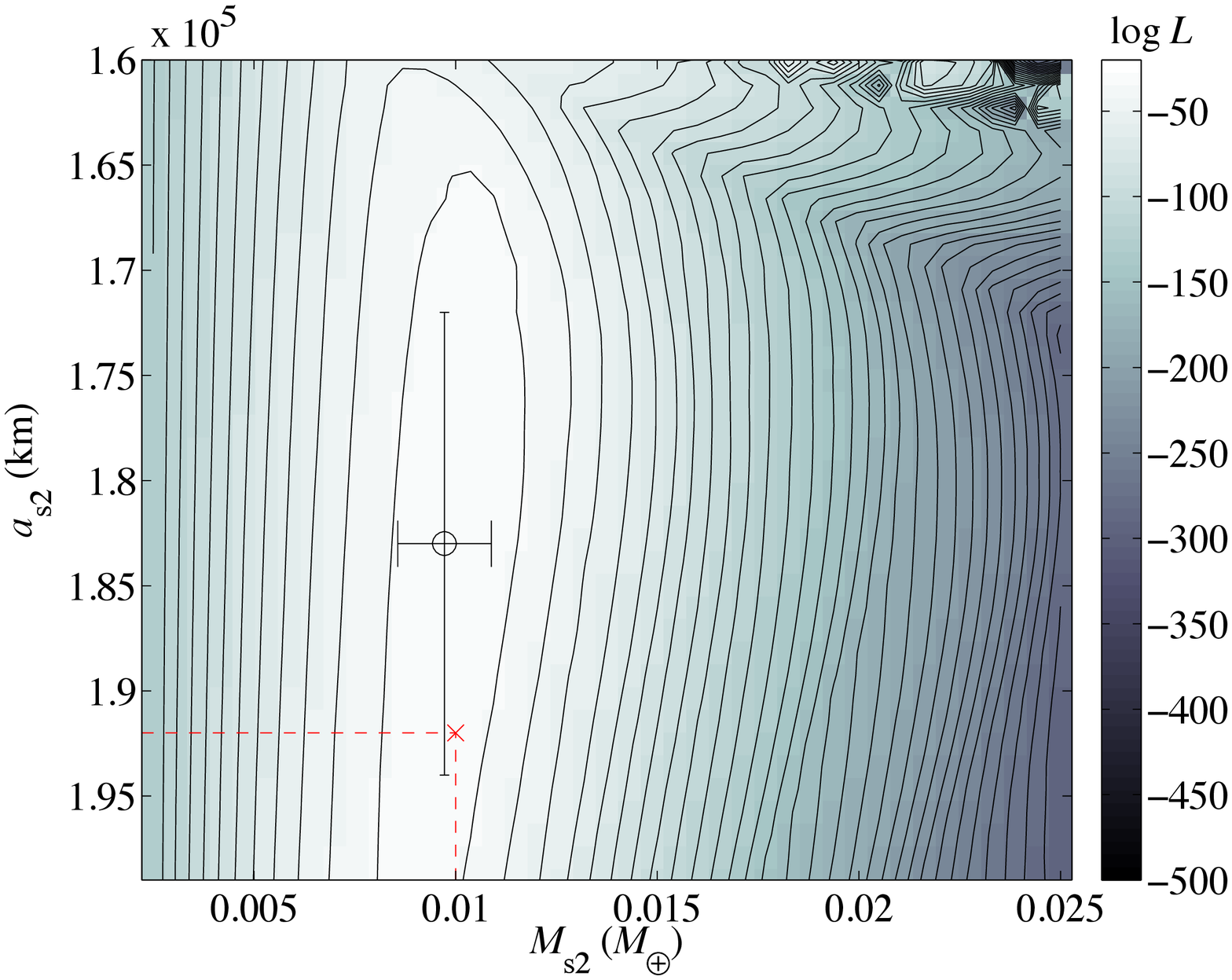}
\caption{\label{fig:probplotsmulti} Log-likelihood landscape of the two-moon model applied to simulated two-moon data. Lighter areas indicate regions of better fits. The red cross indicates the location of the simulated system to be retrieved. The parameters corresponding to the inner moon are shown in the left panel and those of the outer moon are shown in the right panel. The vertical ridge of high log-likelihood in the right panel indicates that the orbital distance of the outer moon ($a_{\rm s2}$) is not as well constrained as that of the inner moon.}
\end{figure*}

The estimates for the two-moon case are shown in Table~\ref{tab:multi-moon}. The favored model in terms of log-evidence is indeed the two-moon model, predicting $M_{\rm s1} = 0.021 \pm 0.002\,M_\oplus$ and $a_{\rm s1} = 131,000 \pm 16,560$\,km. The outer moon is predicted to have $M_{\rm s2} = 0.0097 \pm 0.0012\,M_\oplus$ and $a_{\rm s2} = 183,000 \pm 11,000$\,km. Hence, both pairs of parameters are in good agreement with the true values. The log-likelihood landscape is plotted in Fig.~\ref{fig:probplotsmulti}. The landscape referring to the inner moon (left panel) shows a peak around the true value $(M_{\rm s1},a_{\rm s1}) = (0.02\,M_\oplus, 120,000\,{\rm km})$ rather than the above-mentioned plateau in the one-moon case, implying that the parameters of the inner moon are well constrained. The mass of the outer moon is tightly constrained as well, but $a_{\rm s2}$ has large uncertainties visualized by the high-log-likelihood ridge in the right panel.

\section{Discussion}
\label{sec:discussion}

Our exomoon search algorithm uses an $N$-body simulator to generate TTV-TDV diagrams based on $a_{\rm s}$, $R_{\rm s}$, and optionally $e_{\rm s}$ and $i_{\rm s}$). In search of one-moon systems, however, analytic solutions exist. In particular, the TTV-TDV ellipse of planets hosting one exomoon in a circular orbit can be analytically calculated using the TTV and TDV amplitudes via Eqs.~(3) in \citet{2009MNRAS.392..181K} and (C7) in \citet{2009MNRAS.396.1797K} and multiplying those RMS values by a factor $\sqrt{2}$. This would dramatically decrease the computation times, but computing times are short for our limited parameter range in this example study anyway.

TTVs can also be caused by additional planets \citep{2005MNRAS.359..567A}, but the additional presence or absence of TDVs puts strong constraints on the planet versus moon hypothesis \citep{2012Sci...336.1133N}. Planet-induced TDVs have only been seen as a result of an apsidal precession of eccentric planets caused by perturbations from another planet and in circumbinary planets \citep{2011Sci...333.1602D}. Moreover, an outer planetary perturber causes TTV on a timescale that is longer than the orbital period of the planet under consideration (thereby causing sine-like TTV signals), whereas perturbations from a moon act on a timescale much shorter than the orbital period of the planet (thereby causing noise-like signals).

An important limitation of our method is in the knowledge of $M_{\rm p}$, which is often poorly constrained for transiting planets. It can be measured using TTVs caused by other planets, but a TTV-TDV search aiming at exomoons would try to avoid TTV signals from other planets. Alternatively, stellar radial velocity measurements can reveal the total mass of the planet--moon system. If the moon(s) were sufficiently lightweight it would be possible to approximate the mass of the planet with the mass of the system. Joint mass-radius measurements have now been obtained for several hundred exoplanets. If the mass of the planet remains unknown, then TTV-TDV diagrams can only give an estimate of $M_{\rm s}/M_{\rm p}$ in one-moon systems. As an additional caveat, exomoon searches might need to consider the drift of the pericenters of the moons (Sect.~\ref{sec:twomoons}). This would involve two more parameters beyond those used in our procedure: the initial orientation of the arguments of periapses and the drift rate. This might even result in longterm transit shape variations for sufficiently large moons on eccentric and/or inclined orbits. Finally, another important constraint on the applicability of our method is the implicit assumption that the transit duration is substantially shorter than the orbital period of the moon. For an analytical description see constraints $\alpha3$) (regarding TTV) and $\alpha7$) (regarding TDV) in \citet{2011tepm.book.....K}.

TTV and TDV effects in the planetary transits of a star--planet--moon(s) system are caused by Newtonian dynamics. This dynamical origin enables measurements of the satellite masses and semimajor axes but not of their physical radii. The latter can be obtained from the direct photometric transit signature of the moons. In these cases, the satellite densities can be derived. Direct moon transits could potentially be observed in individual planet-moon transits or in phase-folded transit light curves. A combination of TTV-TDV diagrams with any of these techniques thus offers the possibility of deriving density estimates for exomoons. TTV-TDV diagrams and the orbital sampling effect are sensitive to multiple moon systems, so they would be a natural combination for multiple moon candidate systems.

We also point out an exciting, though extremely challenging, opportunity of studying exomoons on timescales shorter than the orbital period of a moon around a planet. From Earth, it is only possible to measure the angle $\rho$ (or the numerical remainder $r$ of $P_{\rm B}/P_{\rm s}$) in the TTV-TDV diagram (see Fig.~\ref{fig:method}). However, Katja Poppenh\"ager (private communication 2015) pointed out that a second telescope at a sufficiently different angle could observe transits of a given planet-moon system at different orbital phases of the moon. Hence, the orbit of the moon could be sampled on timescales smaller than $P_{\rm s}$. Then $n$ could be determined, enabling a complete measurement of $\rho$, $P_{\rm s}$, and $n$.

Beyond the star--planet--moon systems investigated in this paper, our concept is applicable to eclipsing binary stars with planets in satellite-type (S-type) orbits \citep{1988A&A...191..385R}. Here, the TTV-TDV diagram of the exoplanet host star would provide evidence of the planetary system around it.

\section{Conclusions}
\label{sec:conclusions}

We identify new predictable features in the TTVs and TDVs of exoplanets with moons. First, in exoplanet systems with a single moon on a circular orbit, the remainder of the planet--moon orbital period ($0~{\leq}~r~{\leq}~1$) appears as a constant angle ($\rho$) in the TTV-TDV diagram between consecutive transits. The predictability of $\rho$ determines the relative position of the moon to the planet during transits. This is helpful for targeted transit observations of the system to measure the largest possible planet-moon deflections (at maximum TTV values) or observe planet-moon eclipses (at maximum TDV values).

Second, exoplanets with multiple moons in an orbital MMR exhibit loops in their TTV-TDV diagrams. These loops correspond the reversal of the tangential motion of the planet during its orbital motion in the planet--moon system. We find that the largest integer in the MMR of an exomoon system determines the number of loops in the corresponding TTV-TDV diagram, for example, five loops in a 5:3 MMR. The lowest number is equal to the number of orbits that need to be completed by the outermost moon to produce a closed TTV-TDV figure, for example, three in a 5:3 MMR.

Planetary TTV-TDV figures caused by exomoons are created by dynamical effects. As such, they are methodologically independent from purely photometric methods such as the direct transit signature of moons in individual transits or in phase-folded light curves. Our novel approach can thus be used to independently confirm exomoon candidates detected by their own direct transits, by planet--moon mutual eclipses during stellar transits, the scatter peak method, or the orbital sampling effect.

We performed blind retrievals of two hypothetical exomoon systems from simulated planetary TTV-TDV. We find that an Earth-sized planet with a large moon (10\,\% of the planetary mass, akin to the Pluto-Charon system) around an M dwarf star exhibits TTVs/TDVs that could be detectable in the four years of archival data from the \textit{Kepler} primary mission. The odds of detecting exomoon-induced planetary TTVs and TDVs with \textit{PLATO} are comparable. \textit{PLATO} will observe about ten times as many stars as \textit{Kepler} in total and its targets will be significantly brighter ($4~{\leq}~m_{\rm V}~{\leq}~11$). The resulting gain in signal-to-noise over \textit{Kepler} might, however, be compensated by \textit{PLATO}'s shorter observations of its two long-monitoring fields \citep[two to three years compared to four years of the \textit{Kepler} primary mission;][]{2014ExA....38..249R}. Our blind retrieval of a multiple moon test system shows that the TTV-TDV diagram method works in principle, from a technical perspective. In reality, however, multiple moons are much harder to detect, requiring transit observations over several years by a space-based photometer with a collecting area slightly larger than that of the \textit{Hubble Space Telescope}.

As modern space-based exoplanet missions have duty cycles of a few years at most, exomoon detections via planetary TTVs and TDVs will be most promising if data from different facilities can be combined into long-term datasets. Follow-up observations of planets detected with the \textit{Kepler} primary mission might be attractive for the short term with four years of data being readily available. However, \textit{Kepler} stars are usually faint. Long-term datasets should be obtained for exoplanets transiting bright stars to maximize the odds of an exomoon detection. A compelling opportunity will be a combination of \textit{TESS}, \textit{CHEOPS}, and \textit{PLATO} data. \textit{TESS} (mid-2017 to mid-2019) will be an all-sky survey focusing on exoplanets transiting bright stars. \textit{CHEOPS} (late 2017 to mid-2021) will observe stars known to host planets or planet candidates. Twenty percent of its science observation time will be available for open-time science programs, thereby offering a unique bridge between \textit{TESS} and \textit{PLATO} (2024 to 2030). The key challenge will be in the precise synchronization of those datasets over decades while the timing effects occur on a timescale of minutes.

\vspace{1cm}

\textit{Note added in proof}. After acceptance of this paper, the authors learned that \citet{2012MNRAS.427.2757M} used a TTV-TDV diagram to search for exomoons around WASP-3b. \citet{2013MNRAS.432.2549A} studied the correlation between the squares of the TTV and the TDV amplitudes of exoplanets with one moon.

\clearpage

\clearpage

\clearpage

\clearpage

\clearpage

\begin{appendix}

\section{Higher-order MMRs}
\label{sec:app}

In the following, we present a gallery of TTV-TDV diagrams for planets with up to five moons in a chain of MMRs. This collection is complete in terms of the possible MMRs. In each case, a Jupiter-mass planet around a Sun-like star is assumed. All moon masses are equal to that of Ganymede. The innermost satellite is placed in a circular orbit at an Io-like semimajor axis ($a_{\rm s1}$) with an orbital mean motion $n_{\rm s1}$ around the planet. The outer moons, with orbital mean motions $n_{{\rm s}i}$ (where $2~{\leq}~i~{\leq}~5$), were subsequently placed in orbits $a_{{\rm s}i}=a_{\rm s1}(n_{\rm s1}/n_{{\rm s}i})^{2/3}$ corresponding to the respective MMR.

\begin{figure*}
\includegraphics[width=.5\linewidth]{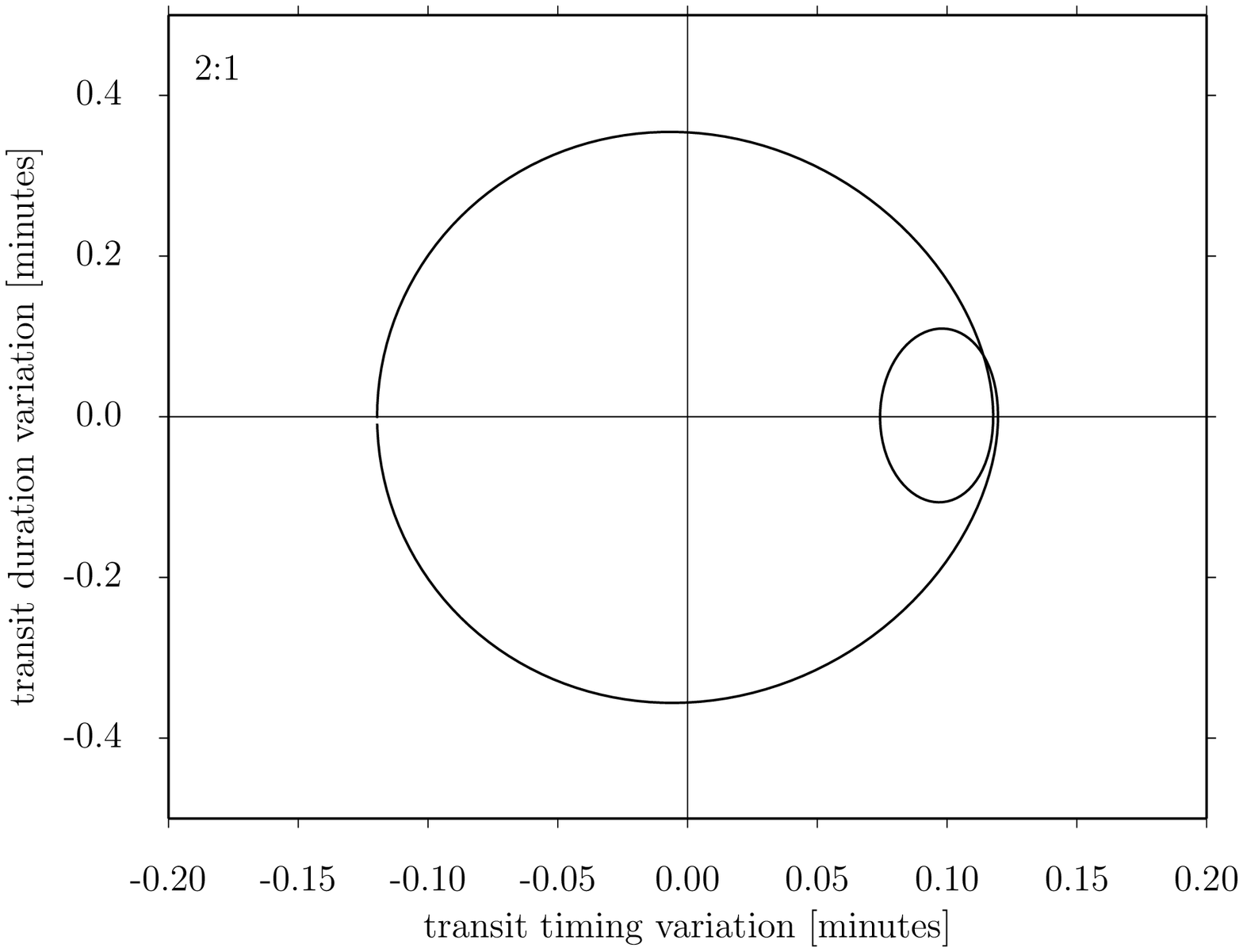}
\includegraphics[width=.5\linewidth]{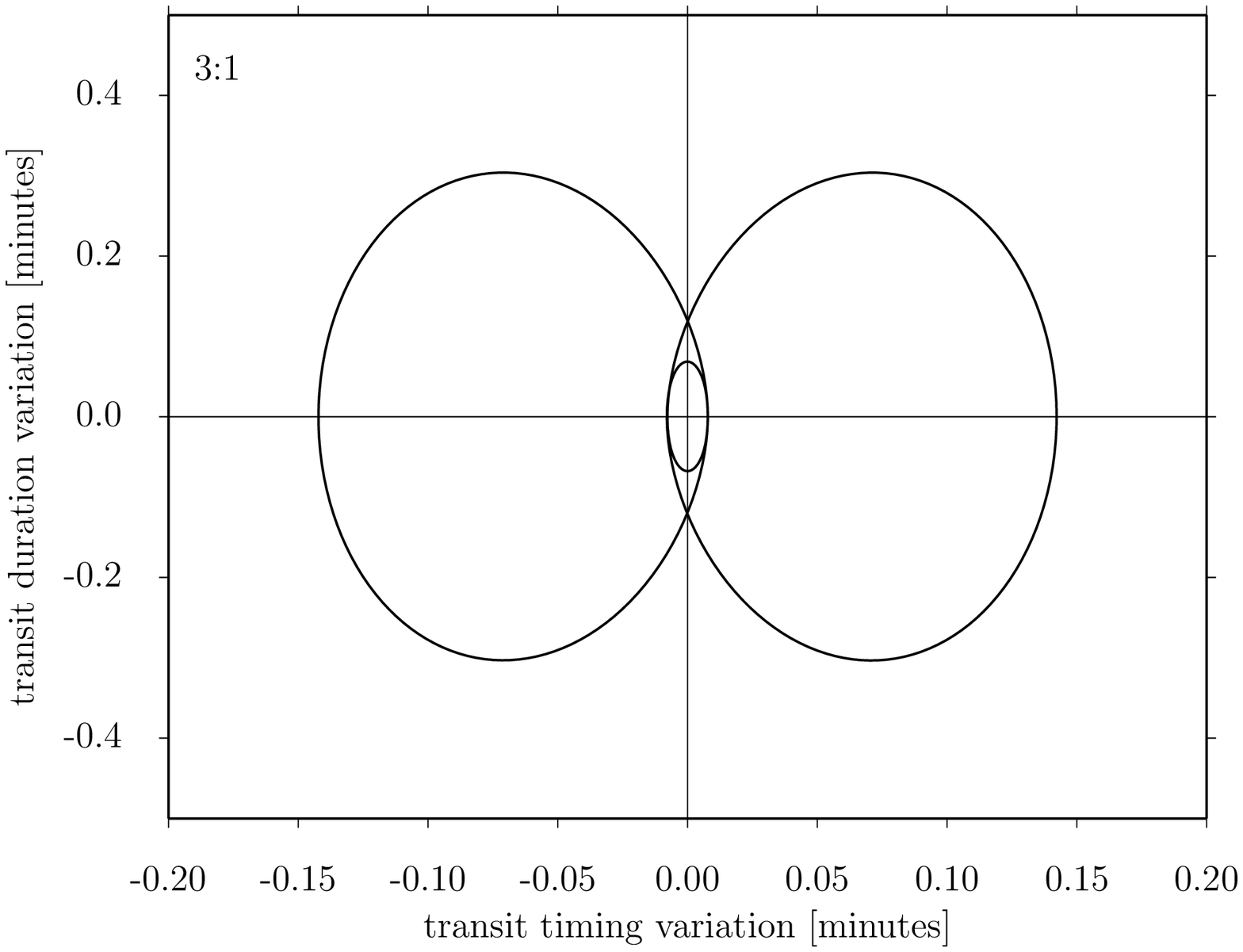}\\
\includegraphics[width=.5\linewidth]{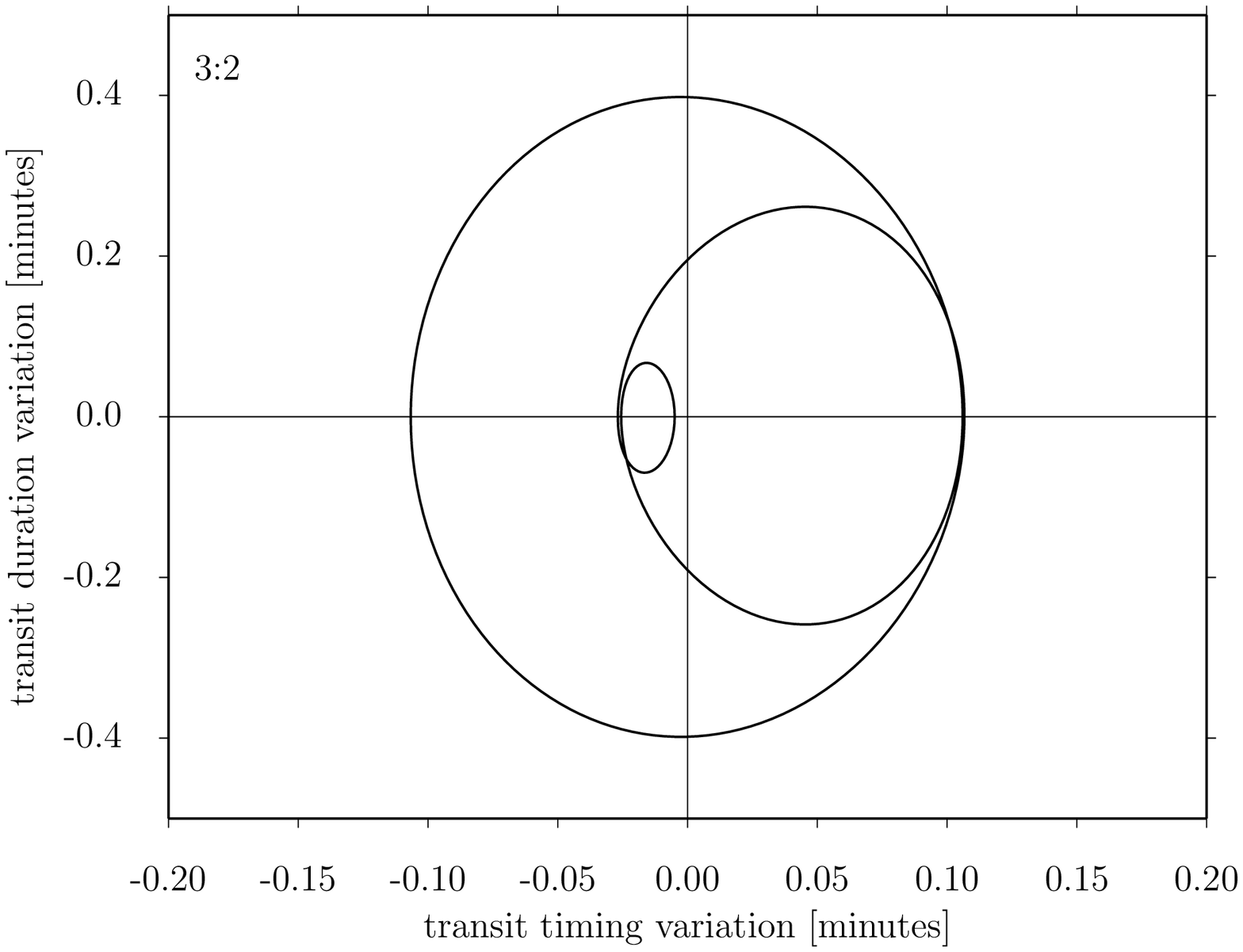}
\includegraphics[width=.5\linewidth]{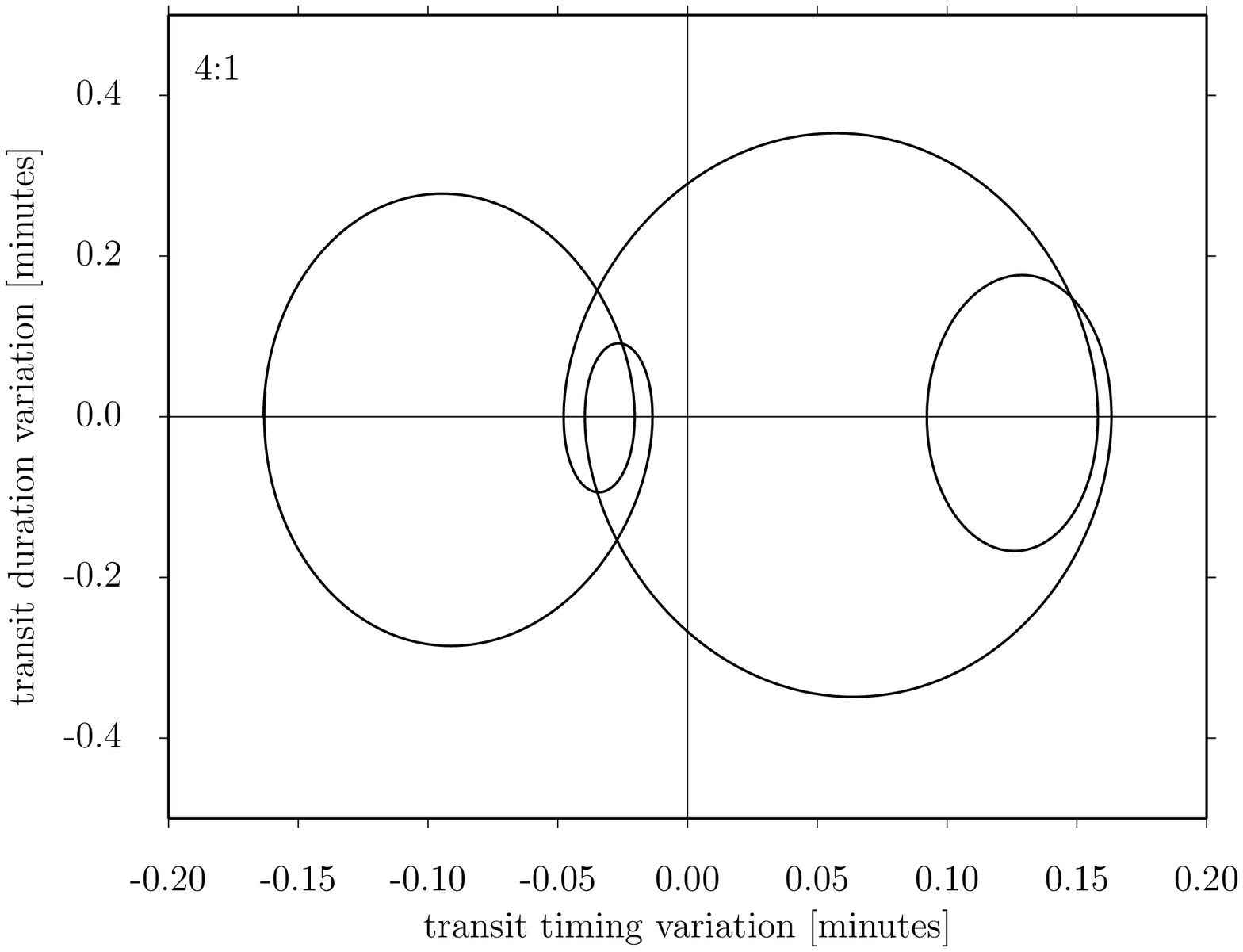}\\
\includegraphics[width=.5\linewidth]{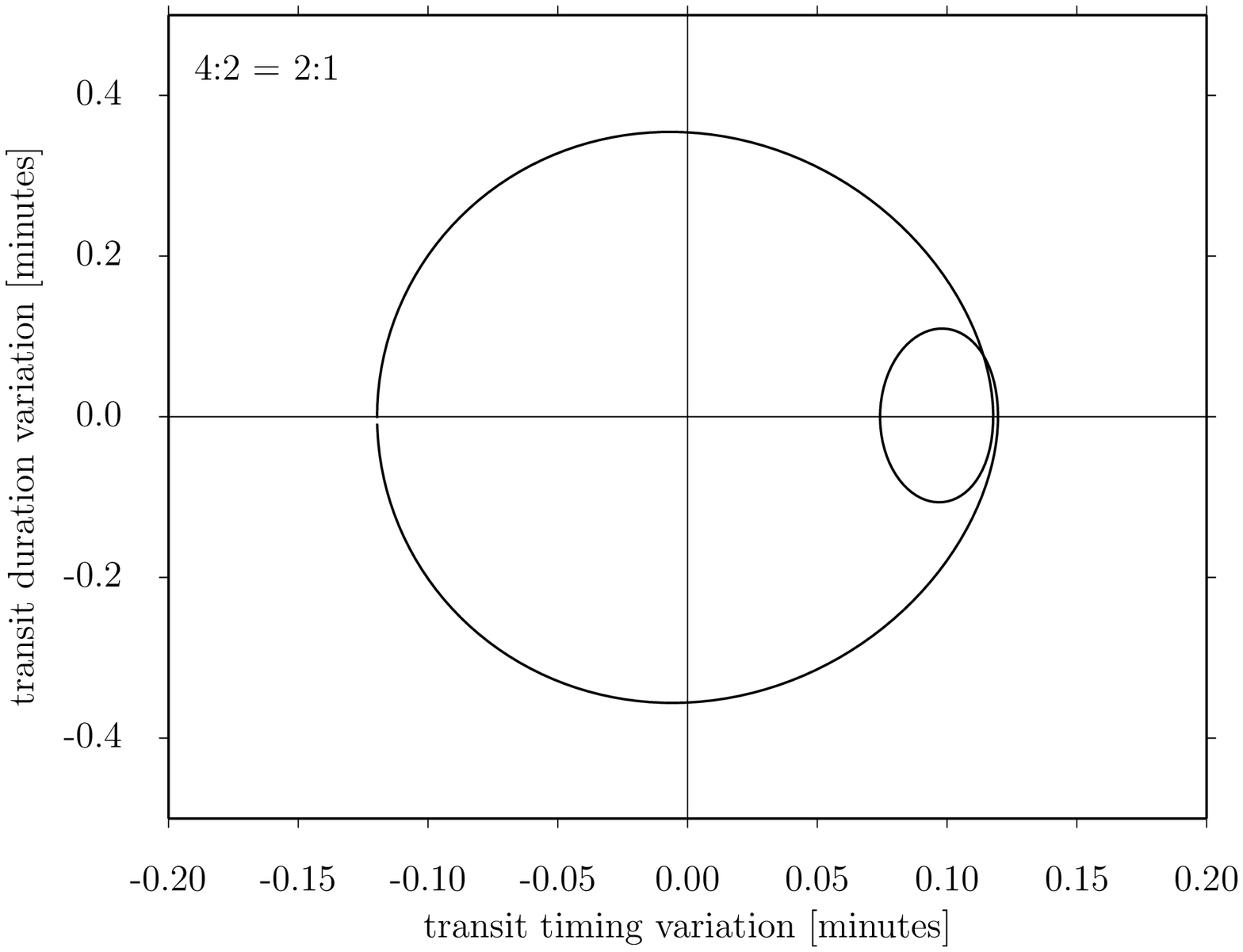}
\includegraphics[width=.5\linewidth]{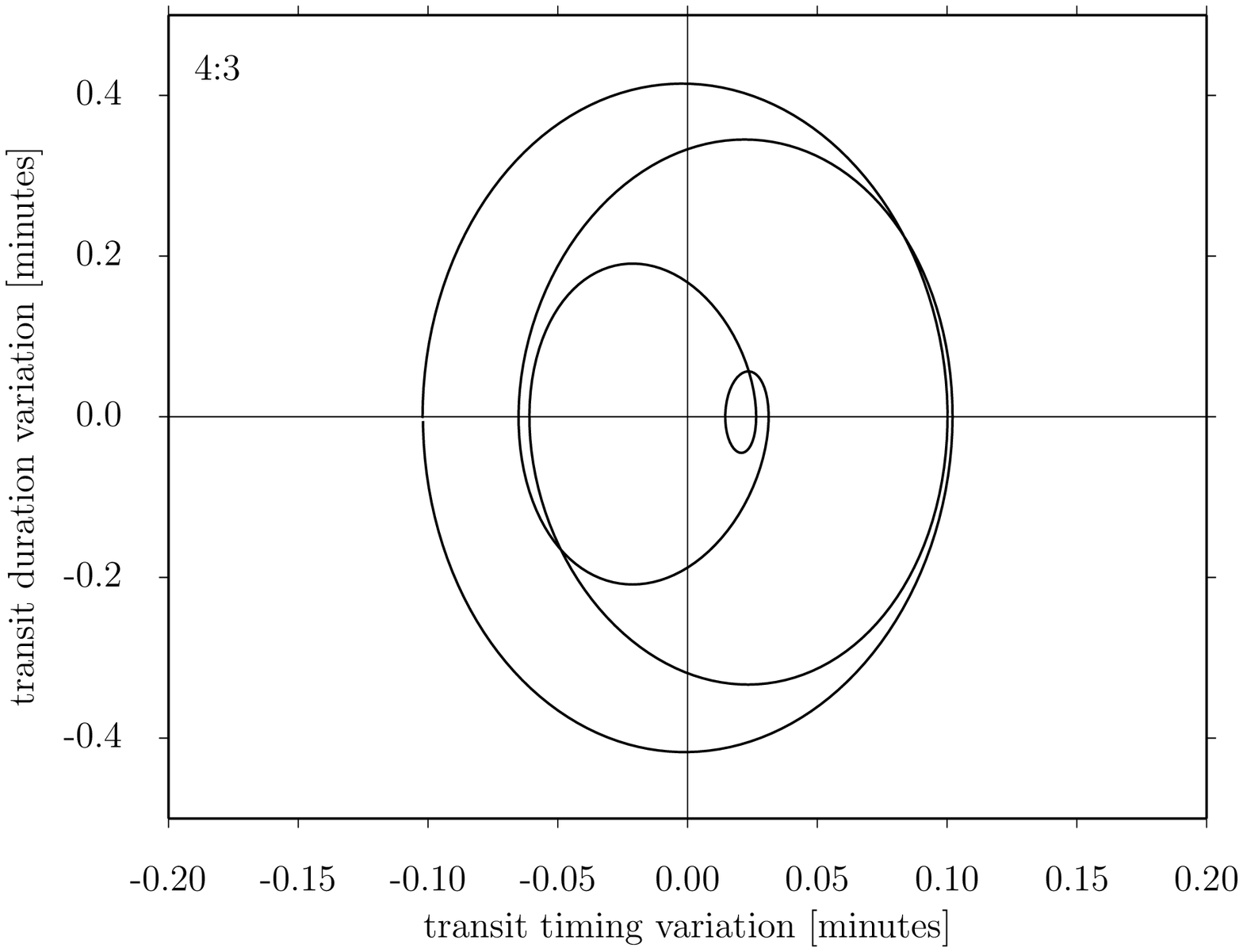}
\caption{\label{fig:higherorder2} TTV-TDV diagrams of planets with two moons in MMRs.}
\end{figure*}

\begin{figure*}
\includegraphics[width=.5\linewidth]{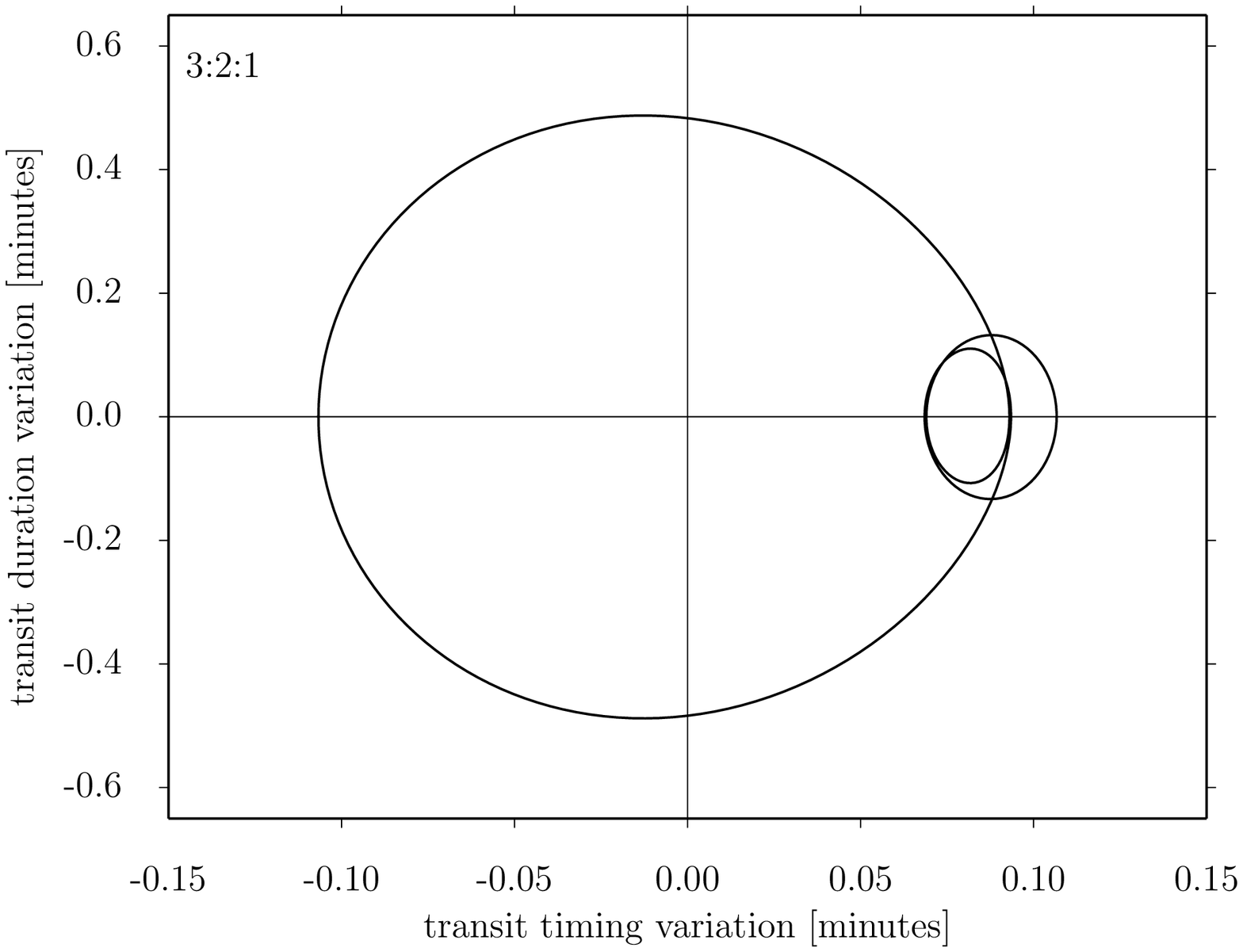}
\includegraphics[width=.5\linewidth]{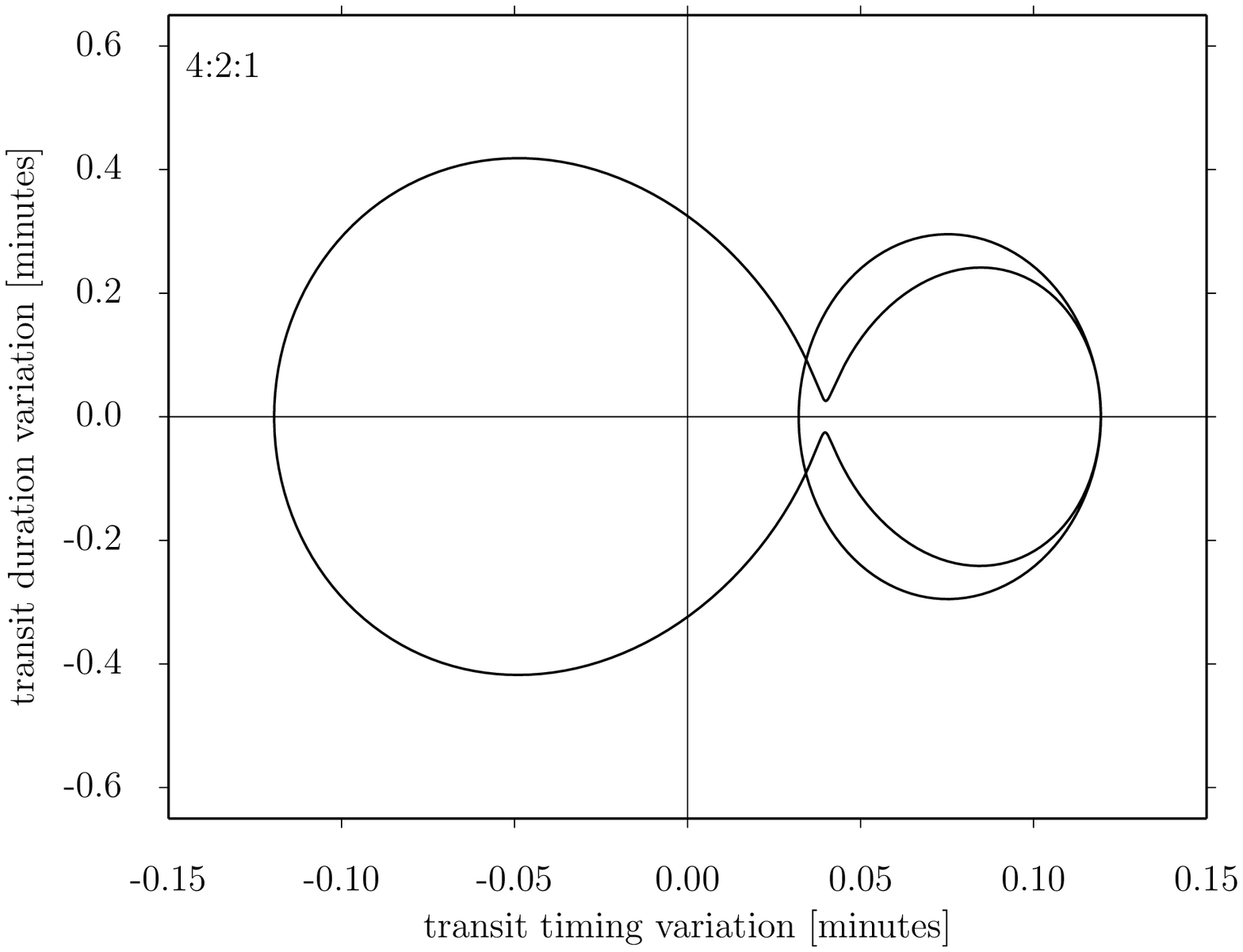}\\
\includegraphics[width=.5\linewidth]{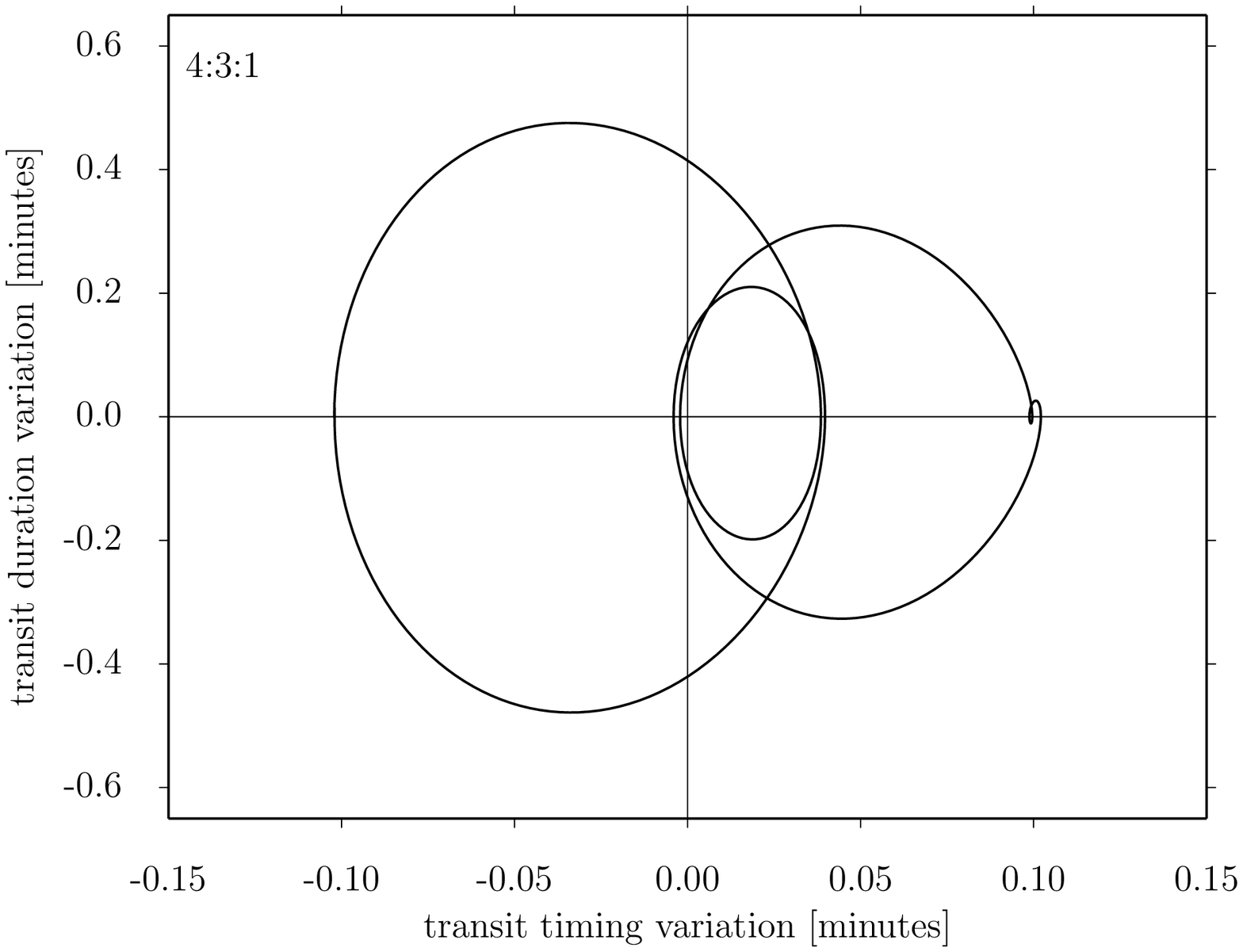}
\includegraphics[width=.5\linewidth]{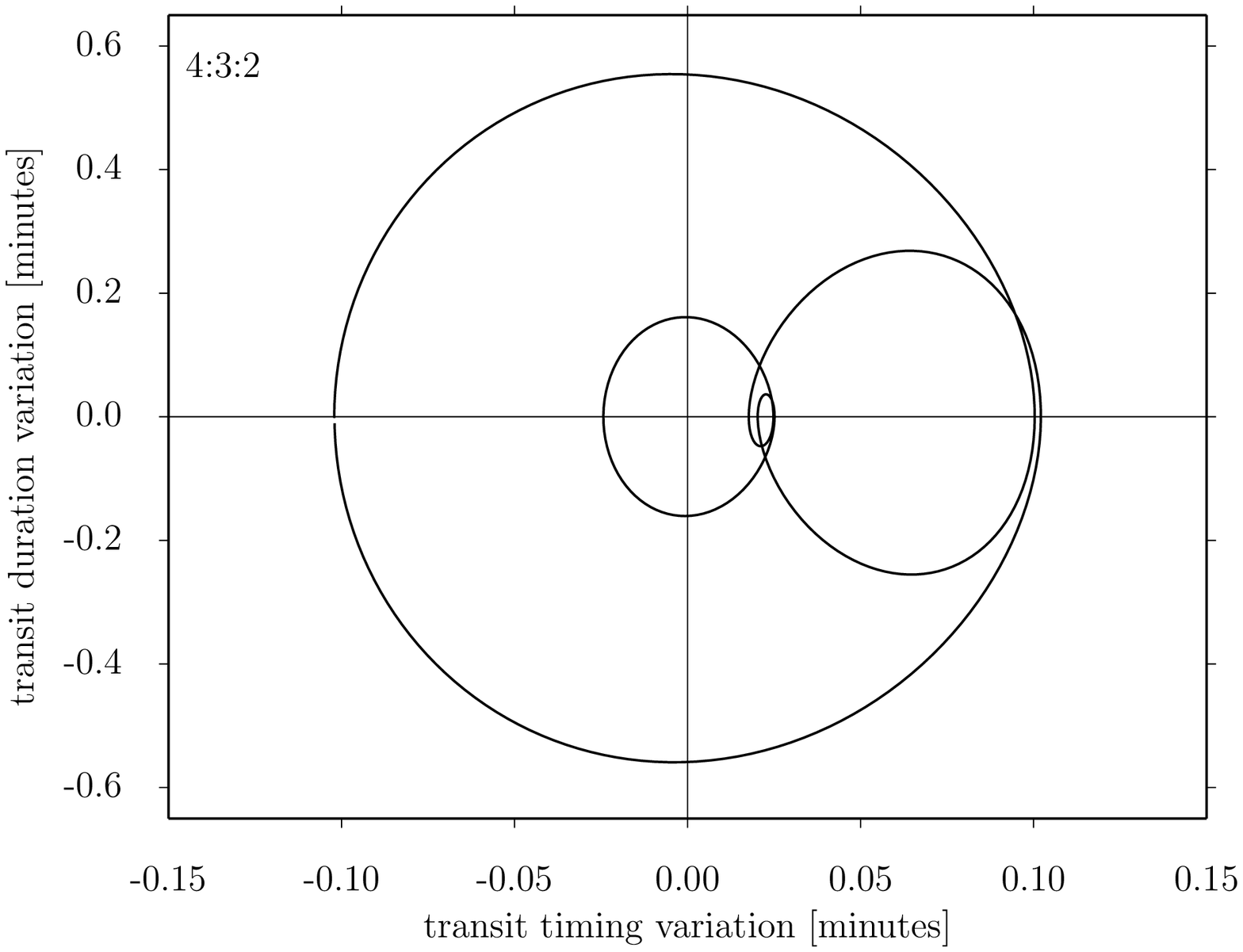}\\
\includegraphics[width=.5\linewidth]{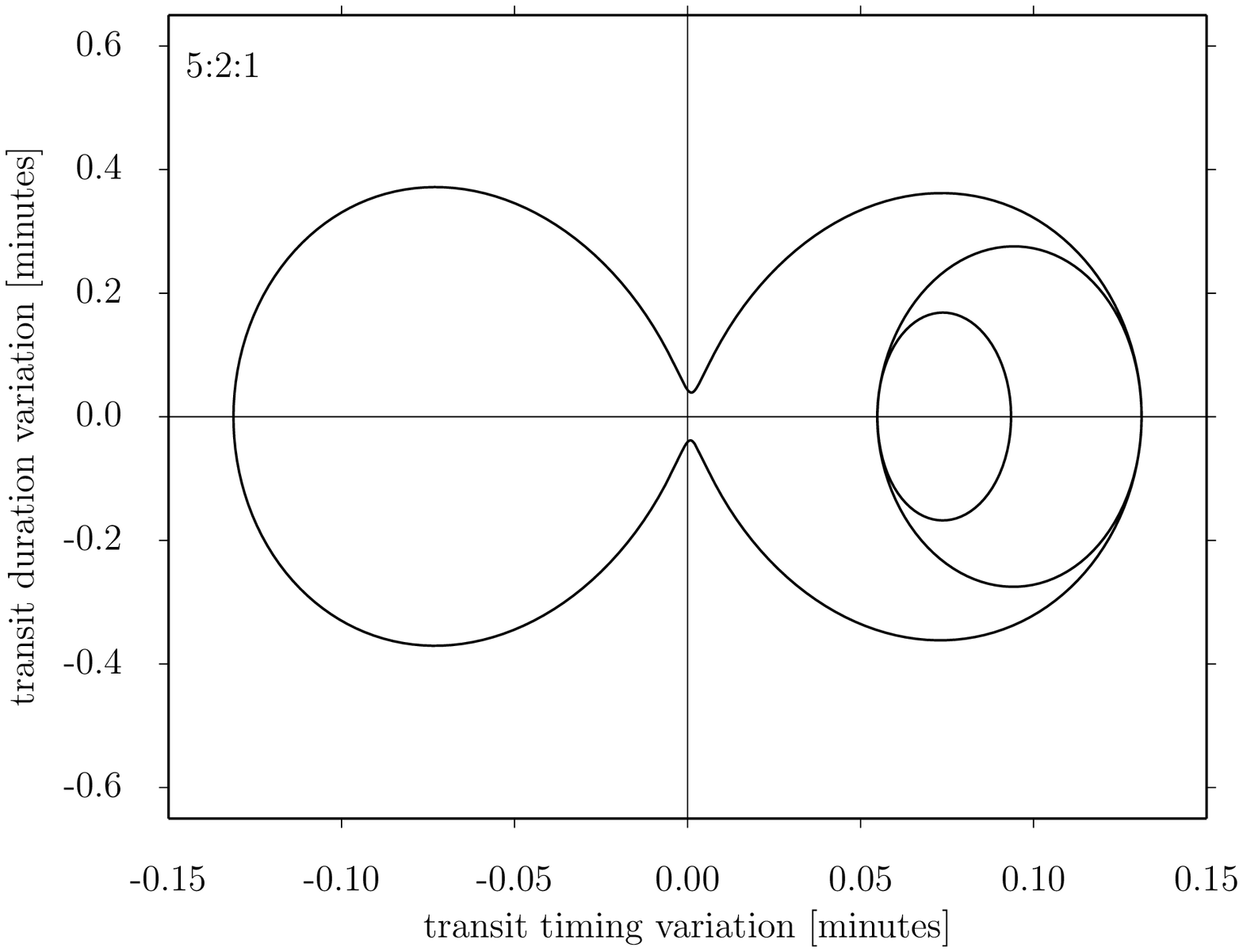}
\includegraphics[width=.5\linewidth]{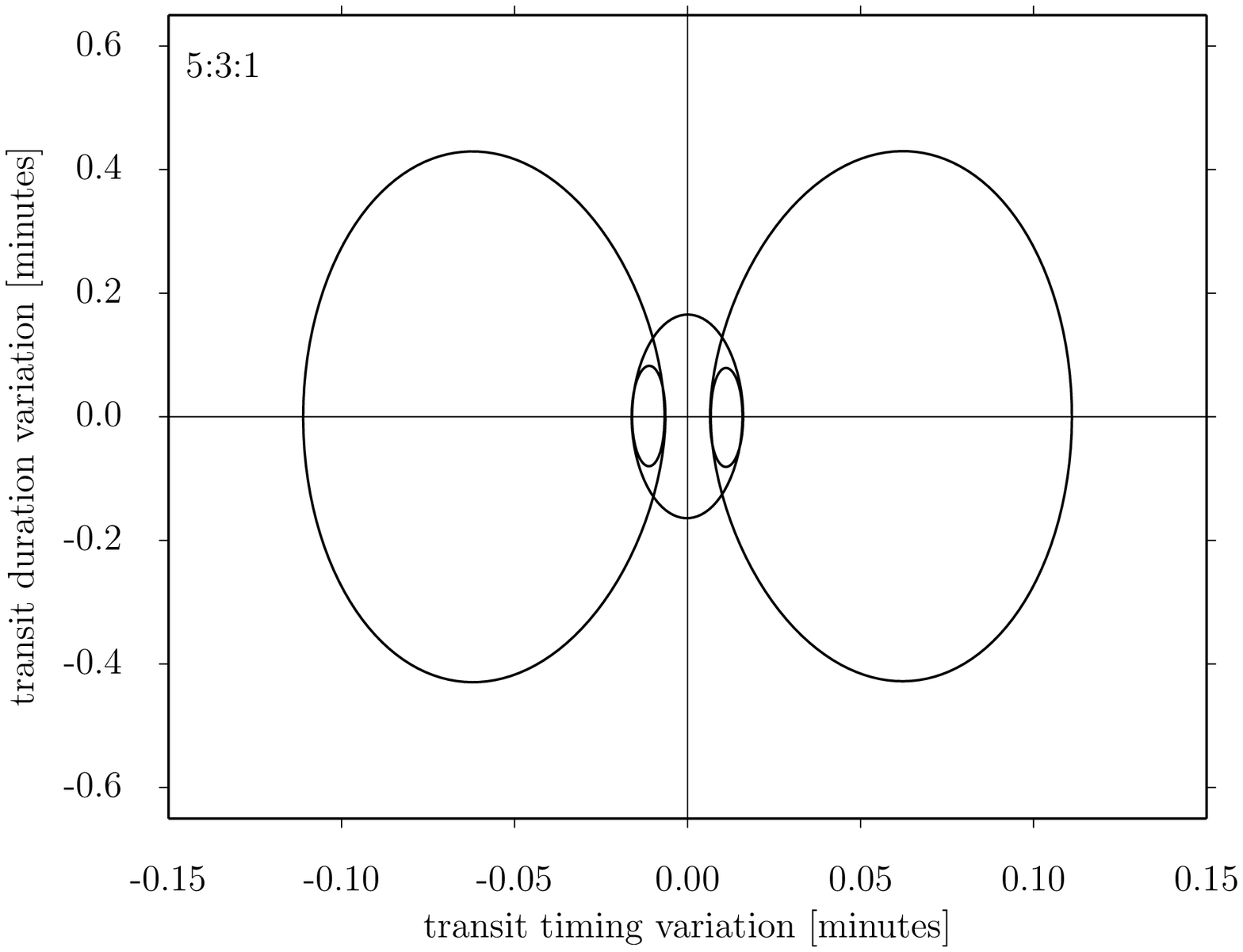}
\caption{\label{fig:higherorder3} TTV-TDV diagrams of planets with three moons in MMRs.}
\end{figure*}

\renewcommand{\thefigure}{A.\arabic{figure}}
\setcounter{figure}{1}

\begin{figure*}
\includegraphics[width=.5\linewidth]{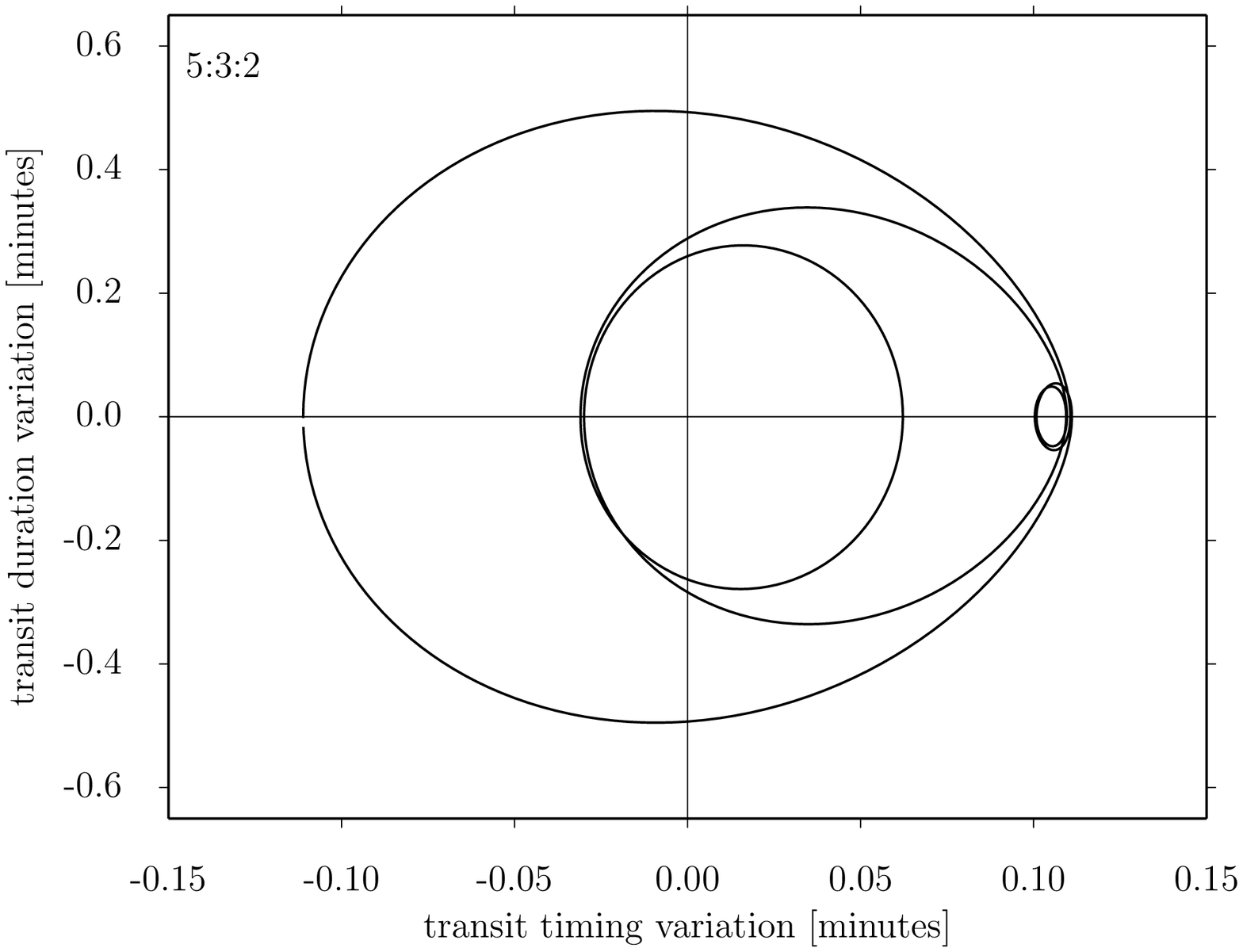}
\includegraphics[width=.5\linewidth]{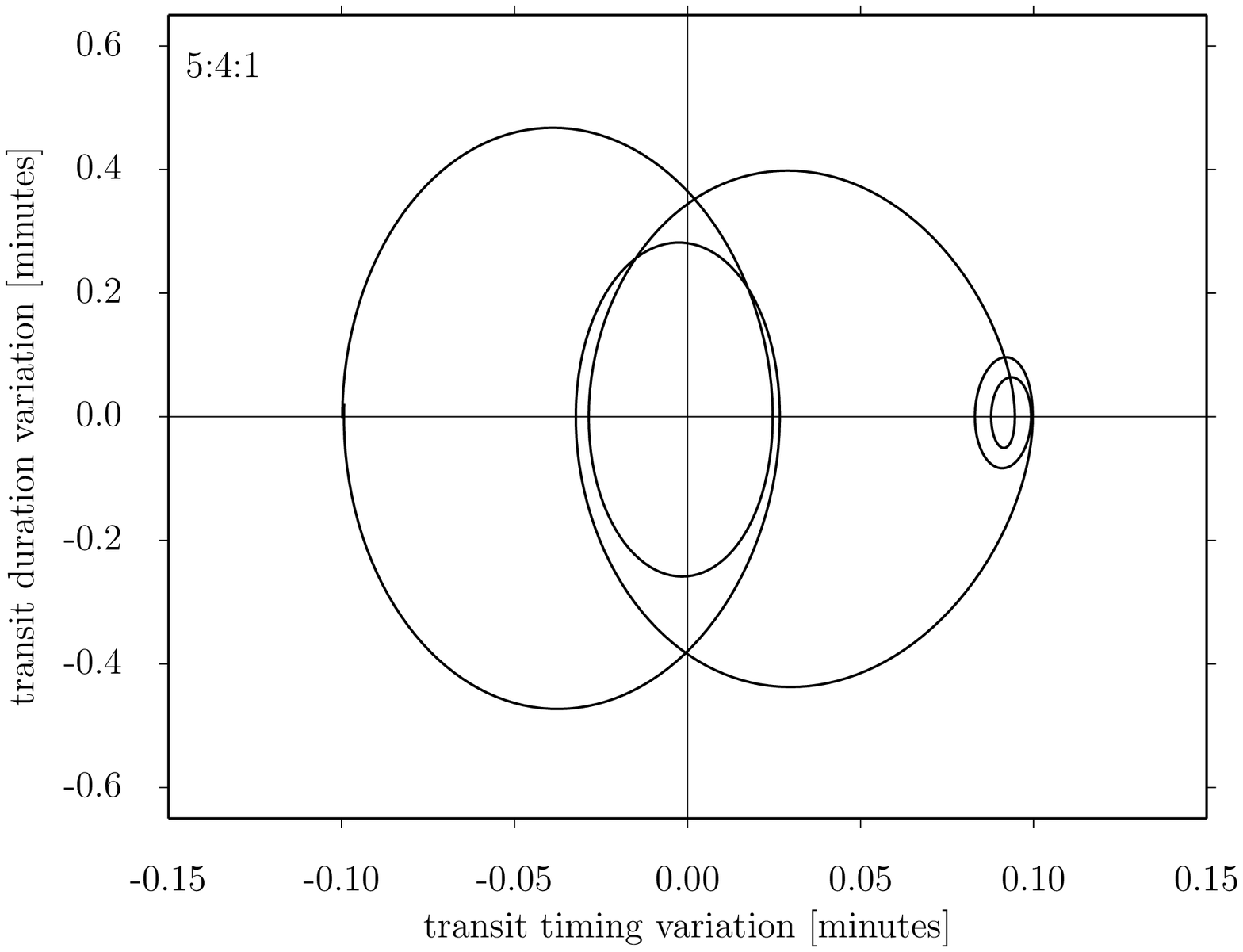}\\
\includegraphics[width=.5\linewidth]{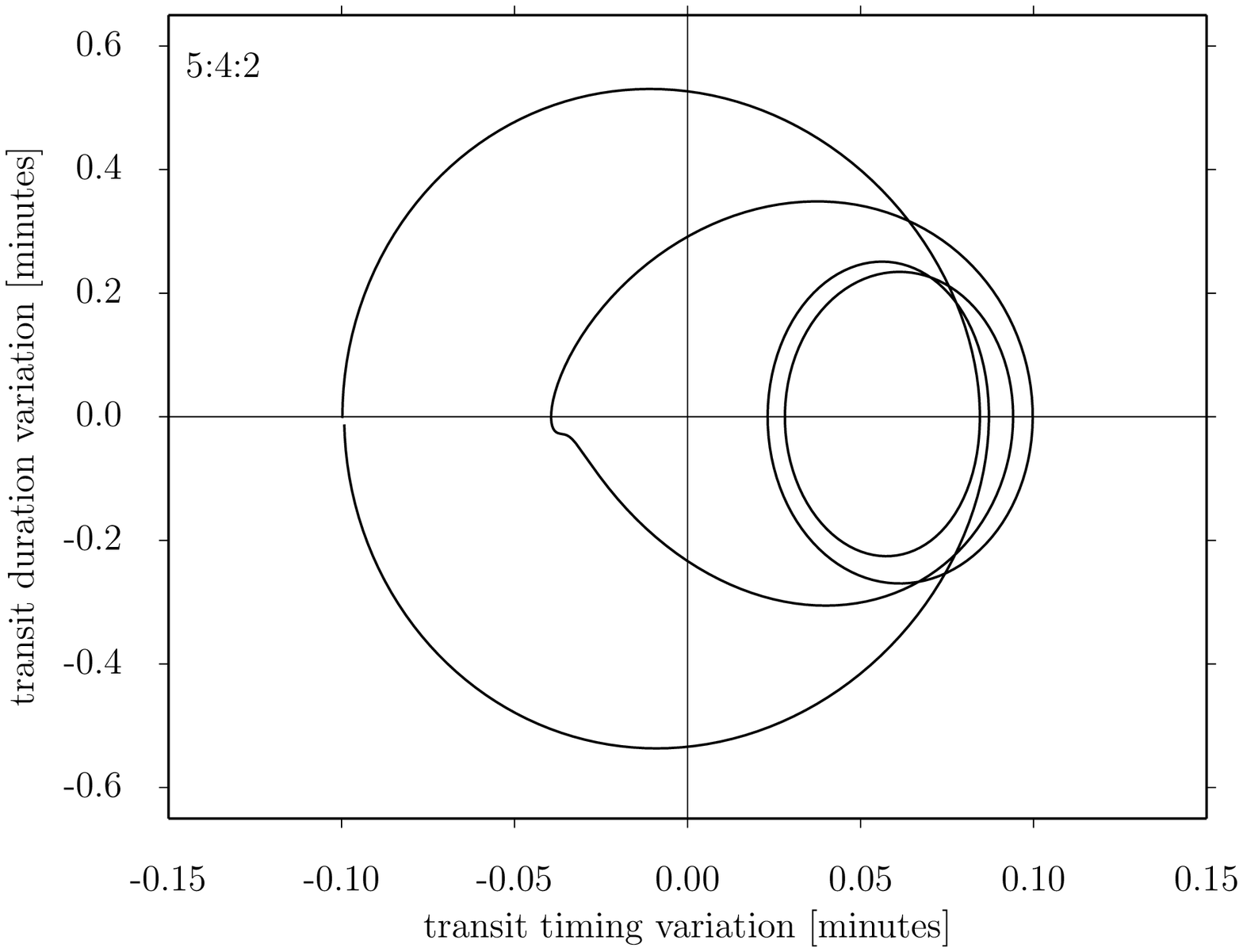}
\includegraphics[width=.5\linewidth]{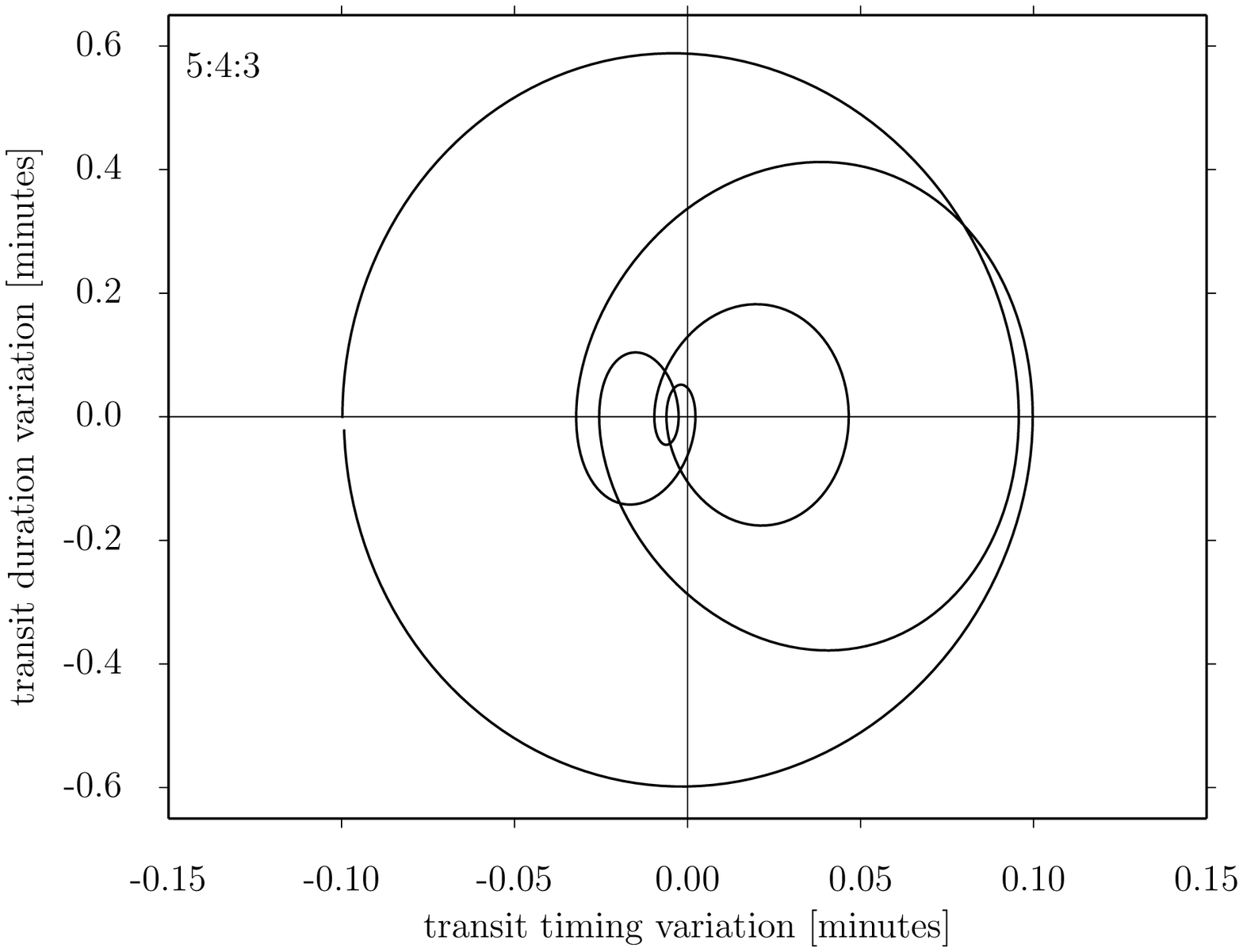}
\caption{(continued)}
\end{figure*}

\begin{figure*}
\includegraphics[width=.5\linewidth]{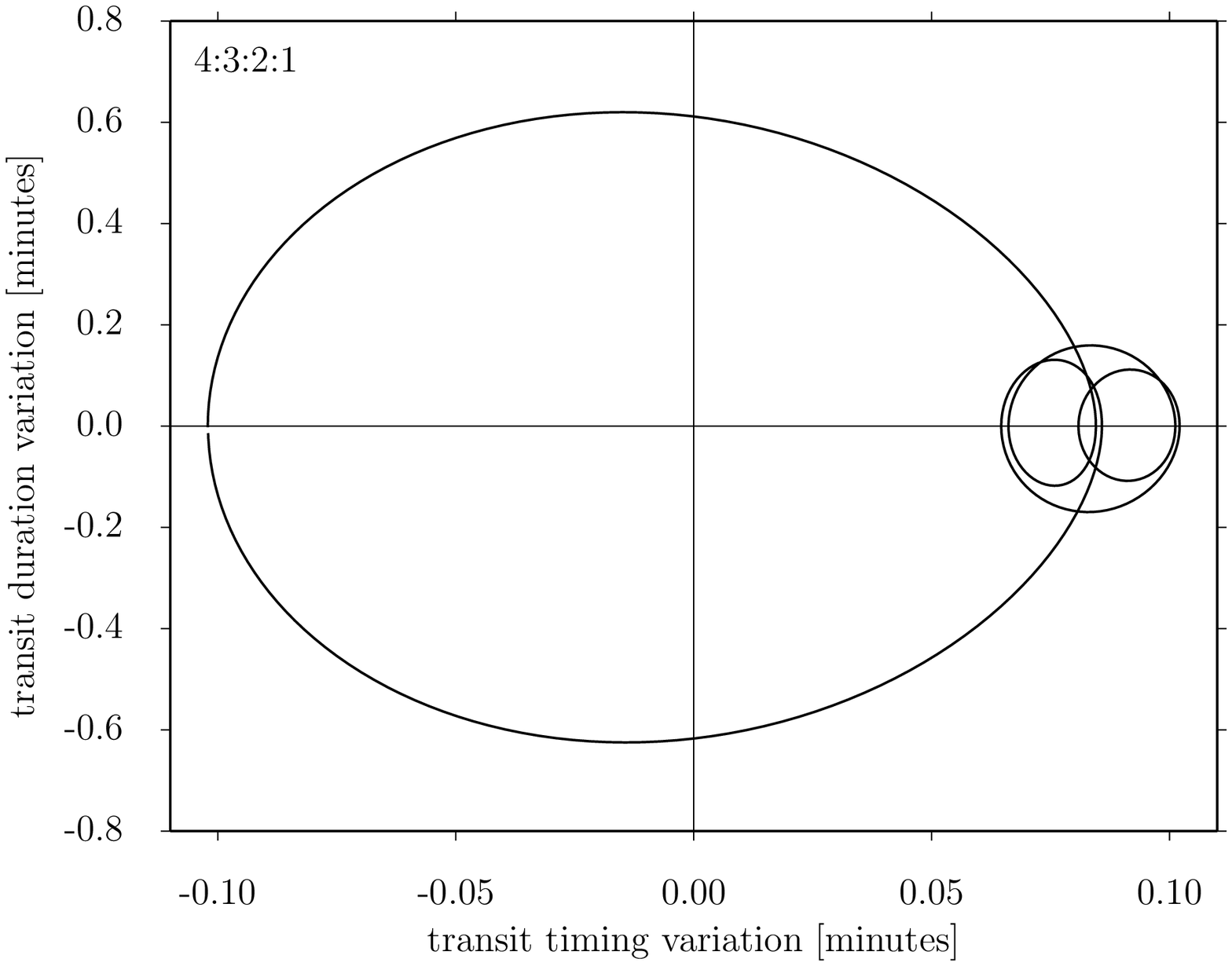}
\includegraphics[width=.5\linewidth]{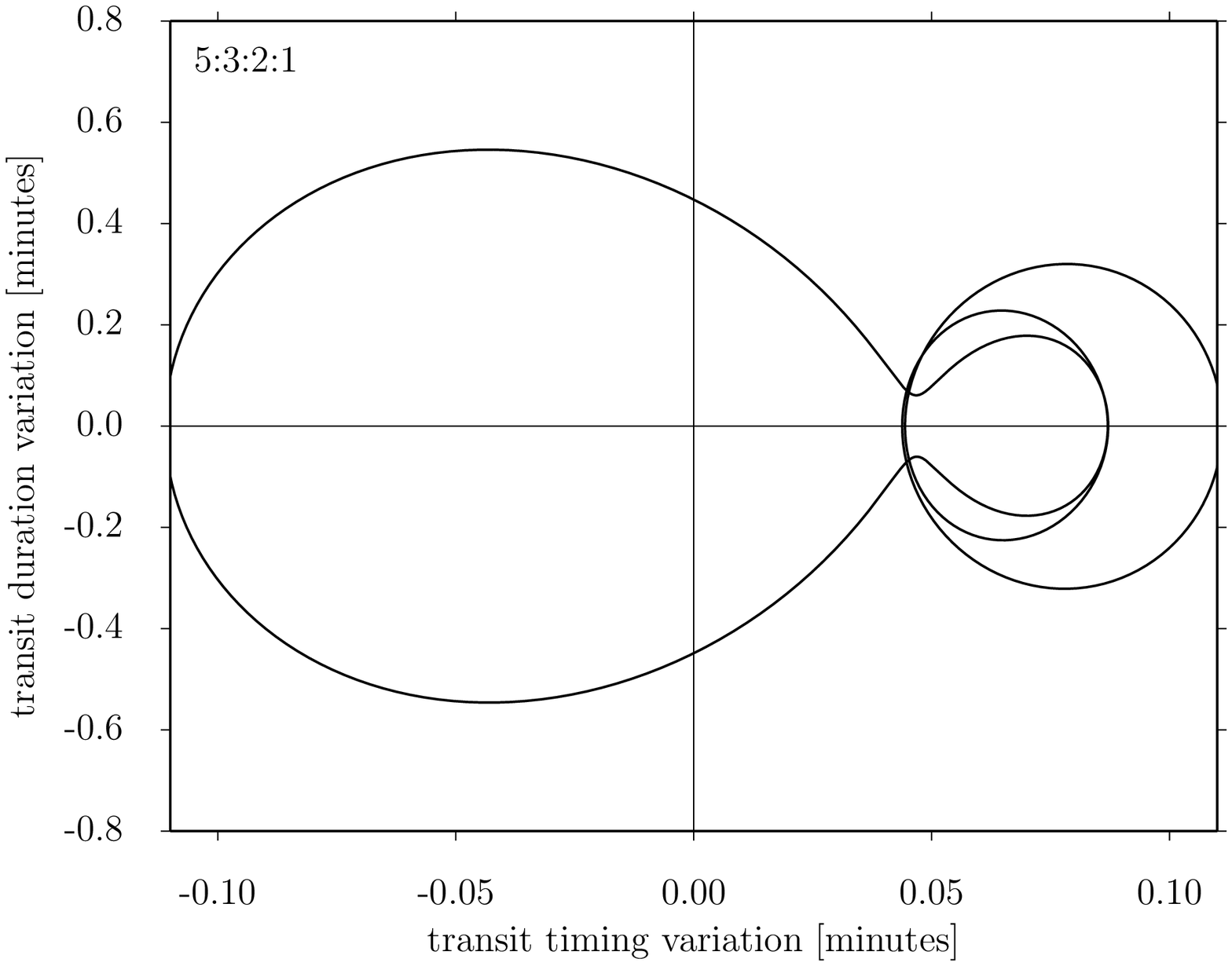}\\
\includegraphics[width=.5\linewidth]{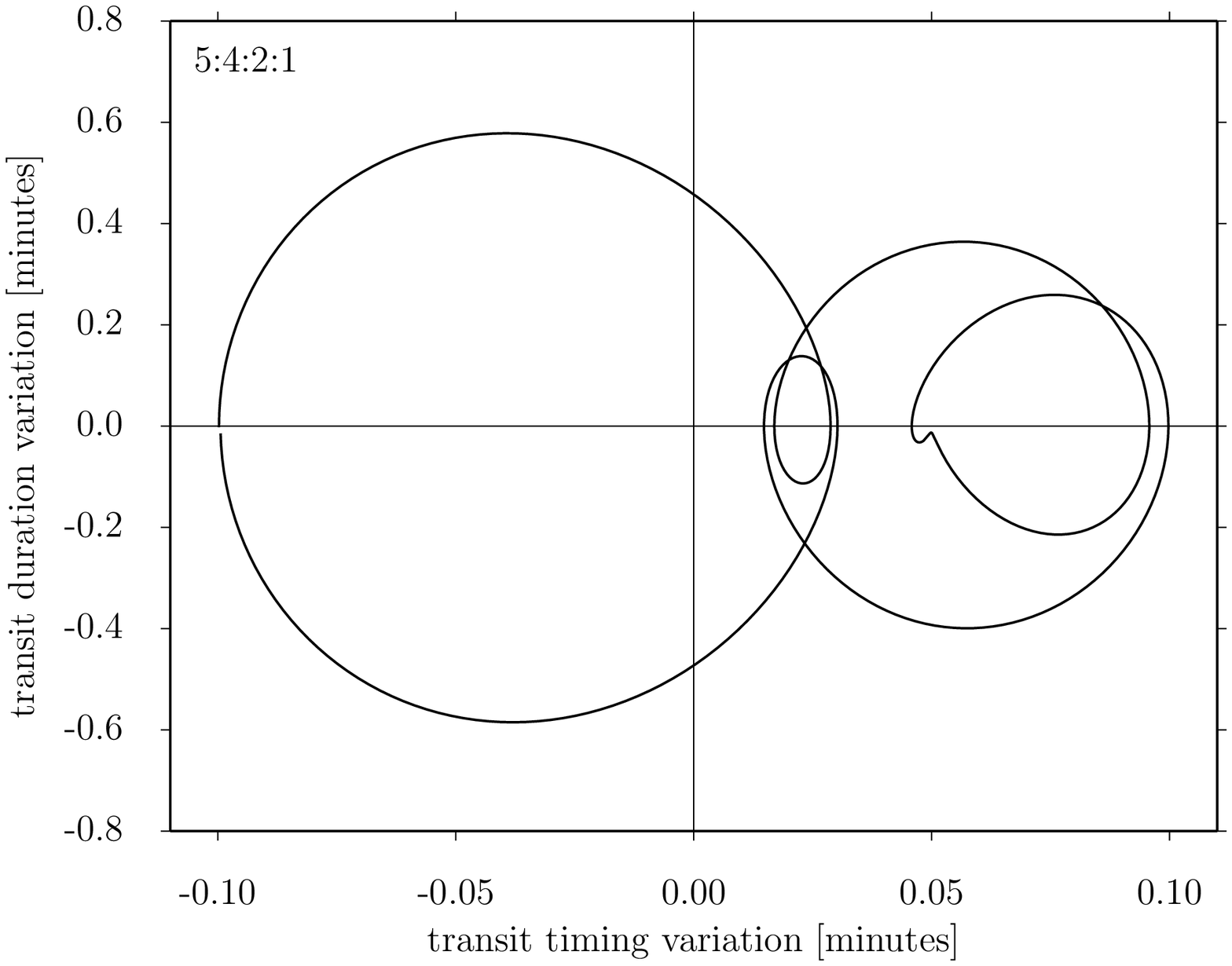}
\includegraphics[width=.5\linewidth]{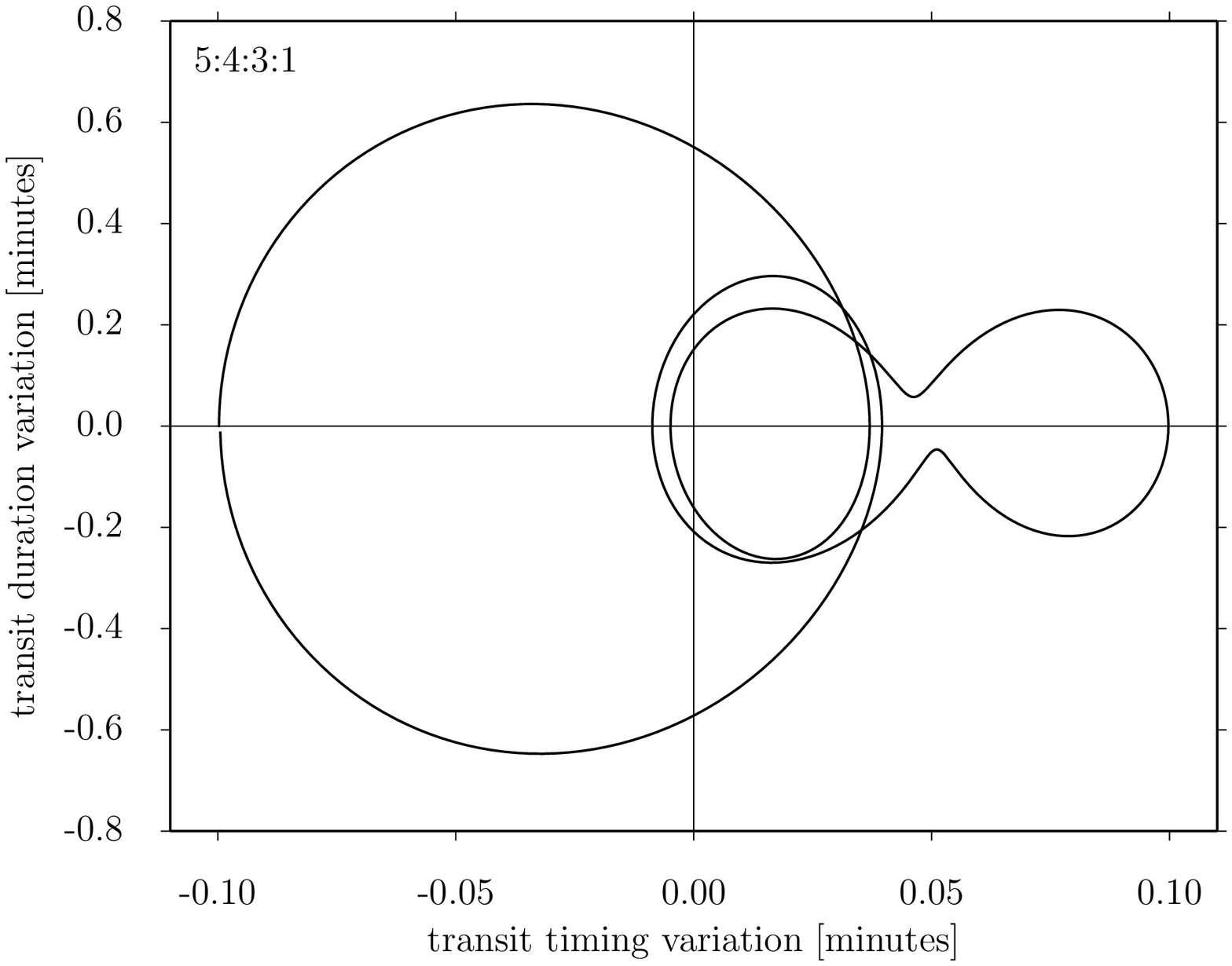}\\
\includegraphics[width=.5\linewidth]{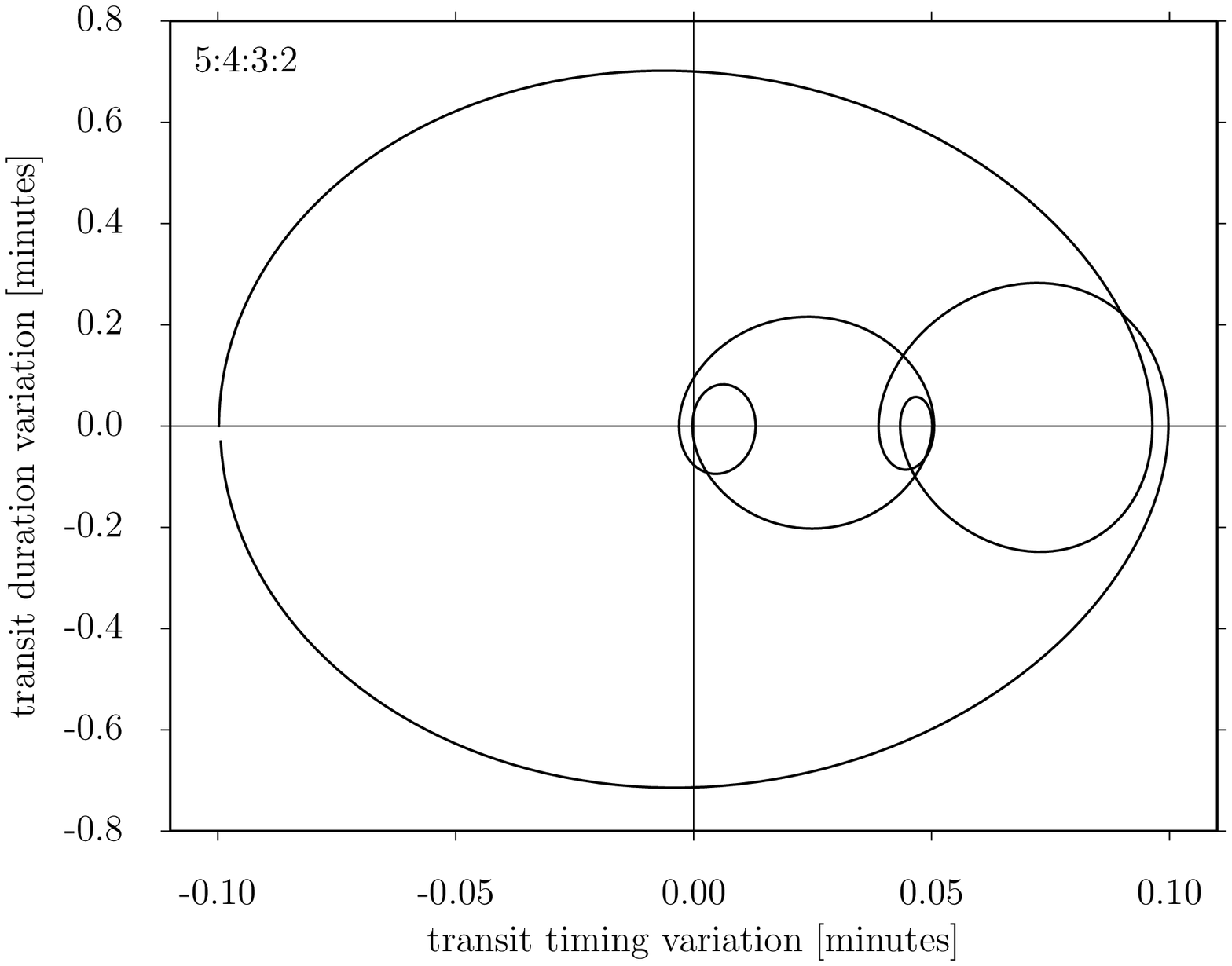}
\caption{\label{fig:higherorder4} TTV-TDV diagrams of planets with four moons in MMRs.}
\end{figure*}

\begin{figure*}
\includegraphics[width=.5\linewidth]{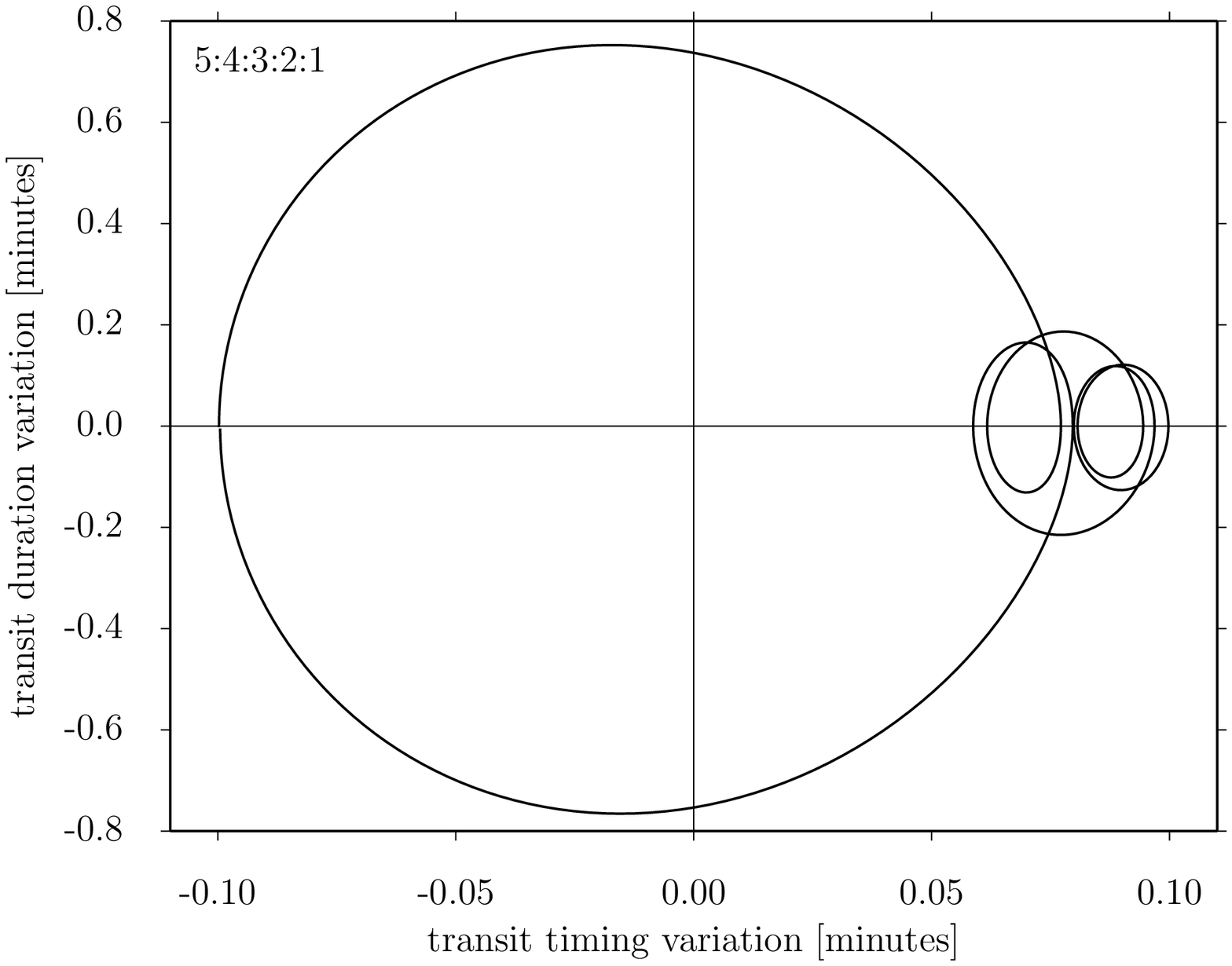}
\caption{\label{fig:higherorder5} TTV-TDV diagram of a planets with five moons in MMRs.}
\end{figure*}

\end{appendix}

\begin{acknowledgements}
We thank Katja Poppenh\"ager for inspiring discussions and the referee for a swift and thorough report. E.~A. acknowledges support from NASA grants NNX13AF20G, NNX13AF62G, and NASA Astrobiology Institute's Virtual Planetary Laboratory, supported by NASA under cooperative agreement NNH05ZDA001C. This work made use of NASA's ADS Bibliographic Services, of the Exoplanet Orbit Database, and of the Exoplanet Data Explorer at \href{www.exoplanets.org}{www.exoplanets.org}.
\end{acknowledgements}


\bibliographystyle{aa} 
\bibliography{ms}

\end{document}